\documentclass[prb,preprint,showpacs]{revtex4}
\usepackage{graphicx}
\usepackage{epstopdf}
\usepackage{amsmath,amssymb}
\usepackage{float}

\newcommand{\ignore}[1]{}
\newcommand{\beq}{\begin{equation}}
\newcommand{\eeq}{\end{equation}}
\newcommand{\mbold}[1]{\mbox{\boldmath $ #1 $}}

\begin{document}

\title
{Effect of the short-range interaction on critical phenomena in 
elastic interaction systems}

\author{Masamichi Nishino$^{1}$}  
\email[Corresponding author. Email address: ]{nishino.masamichi@nims.go.jp} 
\author{Seiji Miyashita$^{2,3}$}

\affiliation{$^{1}${\it Computational Materials Science Center, National Institute}  
for Materials Science, Tsukuba, Ibaraki 305-0047, Japan \\
$^{2}${\it Department of Physics, Graduate School of Science,} The University of Tokyo, Bunkyo-Ku, Tokyo, Japan  \\
$^{3}${\it CREST, JST, 4-1-8 Honcho Kawaguchi, Saitama, 332-0012, Japan}
}
\date{\today}

\begin{abstract} 
The elastic interaction, induced by the lattice distortion due to the difference of the molecular size, causes an effective long-range interaction. 
In spin-crossover (SC) compounds, local bistable states, i.e., high-spin and low-spin states have different molecular sizes, and the elastic interaction is important. 
In bipartite lattices, e.g., the square lattice, the ground state can be two types of phases: ferromagnetic-like and antiferromagnetic-like phases. 
In systems like SC compounds, the former phase consists of all small or large molecules, and the latter phase has the configuration of alternating small and large molecules. In fact, both cases are observed in SC systems. 
In this paper, we have studied the effect of the short-range interaction in the elastic system on the properties of those order-disorder phase transitions. 
We have obtained a phase diagram in the coordinates of the temperature and the strength of the short-range interaction, including the metastable structures. 
We show that effects of the short-range interaction are essentially different for ferromagnetic-like and antiferromagnetic-like phase transitions. 
In the ferromagnetic-like transition, the long-range interaction of elasticity 
is relevant, and the system exhibits a phase transition in the mean-filed universality class. 
In this case, the long-range interaction strongly enhances the ferromagnetic-like order, and it works cooperatively with the short-range interaction. In contrast, in the antiferromagnetic-like transition, the elastic interaction slightly enhances the antiferromagnetic-like order, but essentially it does not contribute to the ordering, and the system shows a transition in the Ising universality class. We have found that in the border region between ferromagnetic-like and antiferromagnetic-like phases, the antiferromagnetic-like phase has an advantage at finite temperatures. 
We discuss the critical properties of two-step SC transitions with comparison between the elastic interaction model and conventional SC models (Ising-like models). 

\end{abstract}


\maketitle

----------------------------------------------------------------------------

\section{Introduction}

Spin crossover (SC) compounds have attracted much attention in a wide variety of phase transitions as well as their potential applications.~\cite{Konig,Hauser,Kahn,Gutlich_book,Letard,Kimura,Pillet,Ichiyanagi,Matsuda,Lorenc,Watanabe,Brefuel,Chong,Bousseksou,Slimani} The spin-crossover system has bistable states, i.e., a low spin (LS) state and a high (HS) state with an entropy difference between the two states.  Entropy-induced phase transition and other cooperative phenomena have frequently been studied by the Ising model with the different degeneracies between the LS and HS states (called Ising-like model) although the Ising interaction was only introduced as the simplest description of the cooperative interaction.\cite{WP}  
From the microscopic viewpoint, the size of each molecule in SC solids changes depending on the spin state, LS or HS, and the importance of the elastic interaction has been suggested for the cooperative interaction.~\cite{Zimmermann,Onishi,Spiering1,Adler,Willenbacher}

By treating the change of the molecular size explicitly, it was shown that the elastic interaction, caused by the lattice distortion due to the difference of the molecular sizes between the LS and HS states, leads to 
the cooperative interaction for the SC phase transition,\cite{Nishino_elastic} where the effective long-range interaction of elastic origin induces a phase transition.  In this limiting case without any short-range interaction, 
the nature of the phase transition has been studied from the view point of the pure long-range interacting model, and it was clarified that the phase transition belongs to the mean-field universality class.\cite{Miya} 
Other important features of the transition have also been clarified with the use of this kind of modeling~\cite{Nishino_elastic,Miya,Konishi,Nishino_elastic2,Enachescu,Enachescu1,Macro_nuc,Enachescu2,SC_book}. 

In realistic compounds, the short-range interaction also plays a role in phase transitions. 
The potential energy function between molecules is considered to depend on the spin states of the molecules.~\cite{Nicolazzi1,Nicolazzi2,Slimani2} 
Then, this dependence is expressed by using a kind of short-range interaction of the spin states. In this case, the interaction between molecules consists of both short-range and long-range components. 

The importance of the short-range antiferromagnetic-like interaction has also been suggested in SC phenomena in the context of explaining two-step transitions.~\cite{Koppen,Petrouleas,Jakobi,Real,Boinnard,Buron} 
An antiferromagnetic-like phase, in which the LS and HS molecules align alternately, is realized as a thermodynamic phase in the middle temperature regions between the LS phase at lower temperatures and the HS phase at higher temperatures. 

In bi-nuclear systems, i.e., two sites in a unit molecule, 
the alternate structure may be easily understood. However, the alternate structure appears also in mono-nuclear compounds, although the crystal has no sublattice symmetry. In theoretical studies of two-step SC transitions, a short-range antiferromagnetic-like interaction was adopted into conventional models such as Ising-like models.~\cite{Bousseksou2,Nishino_two-step,Kamel}. (In Sec.~\ref{sec_Ising-like}, an outline of two-step SC transitions and a related discussion are given.)

Competition and interplay between the short-range and long-range interactions are interesting topics in phase transitions.\cite{Pluis,Low,Sagui,Muratov,Sciortino,Giuliani,Nakada1,Nakada2} In the pure short-range model, clustering of the ordered phase takes place near the critical temperature, leading to the divergence of the correlation length of the order parameter. However, the long-range interaction suppresses the generation of domains, and the configuration is uniform even at the critical temperature.\cite{Miya} 
The crossover from the short-range Ising universality class to the mean-field universality class was studied for the case of a ferromagnetic short-range interaction.\cite{Nakada1,Nakada2} 
In those works, it was found that the long-range interaction is always relevant, and the system exhibits a phase transition in the mean-field universality class.

Elastic interactions with a size difference between atoms have been also studied for alloy systems~\cite{Dunweg,Laradji,Zhu}, where elastic potential energies with empirical parameters were adopted for neighboring atomic species (Si-Si, Si-Ge, Ge-Ge), and a bond angle potential was introduced to maintain the diamond structure. In those models different elastic constants for different atomic species lead to a kind of short-range interactions. 
The critical properties were found to be mean-field like for ferromagnetic-like order, while they were suggested to be of rigid Ising type for antiferromagnetic-like order, although the transition temperature is significantly different from that of the Ising model. 
However, for the ferromagnetic-like order, the elastic interaction of these models causes asymmetry of the entropy effect between Si-Si and Ge-Ge phases due to different elastic constants for different atomic species. Asymmetry of the entropy effect between two phases is also induced by the bond angle potential as we discuss in Appendix~\ref{appendix_angle}. 
Indeed, in those studies an artificial field was applied to avoid the asymmetry and to constrain the systems to the coexistence line.  

In the present work, we use the same elastic constant for different species. 
 To maintain the lattice structure (square lattice), we adopt next nearest neighbor interactions, in which the asymmetry of the entropy effect is negligible as we see in Appendix~\ref{appendix_angle}.  We focus on the dependence of the critical phenomena on the parameters of the short-range interaction. 
Our modeling enables us to study systematically the effect of the short-range interaction on the critical phenomena without applying an additional field to constrain the systems to the coexistence line. Here we can avoid not only the ambiguity and complexity of the asymmetry of the entropy effect but also contamination of the short-range interactions due to different elastic constants for different molecular species.

In this paper, we systematically investigate the effects of both ferromagnetic-like and antiferromagnetic-like short-range interactions on the ordering process in a unified model and present a phase diagram in the coordinates of the temperature and the strength of the short-range interaction, including the metastable structures.  
We find that the nature of the critical phenomena is different in the cases of ferromagnetic-like and antiferromagnetic-like transitions. 
We clarify that the contributions of the short-range interaction and the elastic interaction to enhancement of the ordering are essentially different in the ferromagnetic-like and antiferromagnetic-like transitions. 
We investigate in detail the border region between ferromagnetic-like and antiferromagnetic-like phases and find that the antiferromagnetic-like phase has an advantage at nonzero temperatures. 
To expand also discussion of the critical properties of two-step SC transitions, 
we briefly summarize the characteristic features of the types of SC transitions  from the view point of the phase diagrams of the Ising-like models 
and examine the difference of the properties between the Ising-like models and the elastic interaction model. 
We also analyze several kinds of interactions to maintain the lattice structure (symmetry), focusing on asymmetric entropy effects for broken symmetry phases due to the difference of molecular sizes.

The rest of the paper is organized as follows. 
In Sec.~\ref{sec2}, the model and method are presented.
In Sec.~\ref{secthir}, we discuss the critical properties of ferromagnetic-like and antiferromagnetic-like order parameters.  
In Sec.~\ref{sec_phase_diagram}, we show the phase diagram of the present model in the coordinates of temperature $T$ vs. the short-range interaction parameter $J$. 
In Sec.~\ref{sec_Ising-like}, we show an outline of two-step SC transitions. 
In Sec.~\ref{summary}, we discuss the critical properties of two-step SC transitions and give summary and discussion. 
In Appendix~\ref{appendix_angle}, we give a discussion about types of interactions to maintain the square lattice and about the symmetry between HS and LS. 

\begin{figure}[h]
\centerline{\includegraphics[clip,width=11cm]{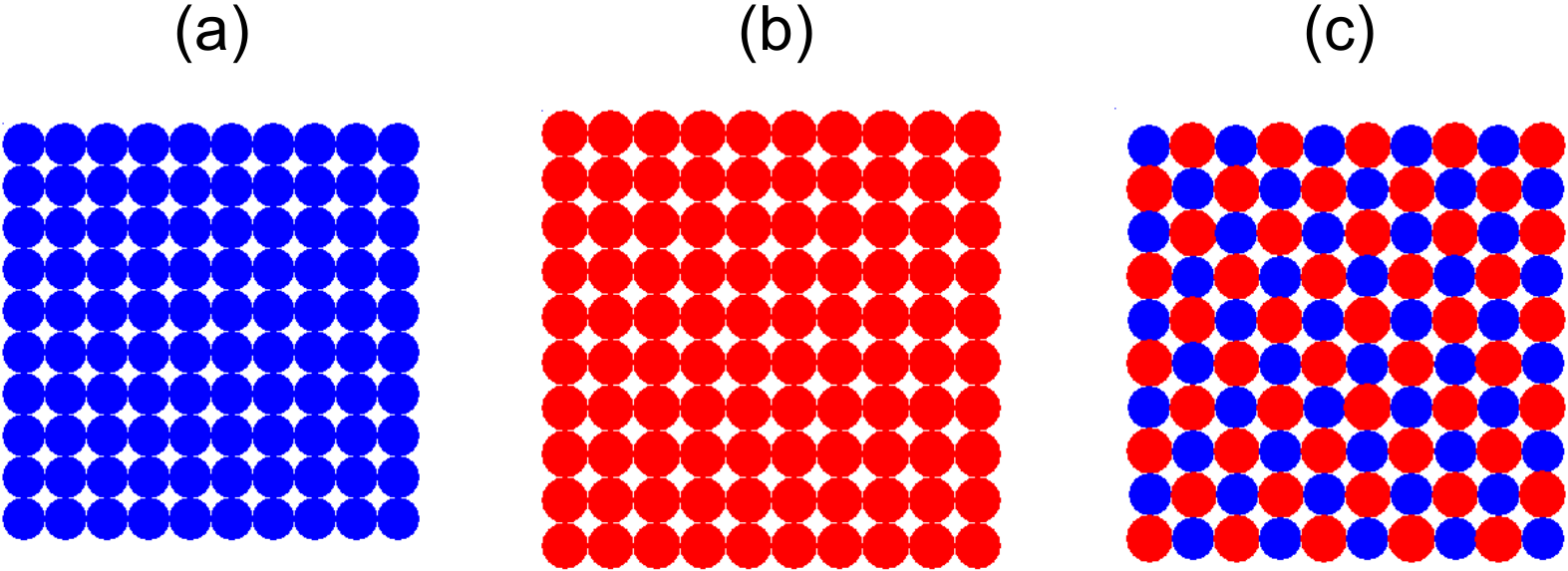} 
}
\caption{ (color online). (
a) Configuration of the ferromagnetic-like phase (low-spin phase), 
(b) Configuration of the ferromagnetic-like phase (high-spin phase), 
(c) Configuration of the antiferromagnetic-like phase. 
Blue and red denote low-spin and high-spin states, respectively. 
High-spin molecules are larger than low-spin molecules. 
}
\label{Fig_conf_T0}
\end{figure}

\section{Model and method}
\label{sec2} 

We consider a system which consists of molecules on a square lattice. 
The positions of the molecules $\{\mbold{r}_i\}$ and also their radii can be changed, and then the lattice can be distorted.  We assume that the square-lattice topology is not broken. 
We adopt the following elastic interactions between the nearest neighbor molecules, 
\begin{equation}
{\cal H}_{\rm nn}={k_1 \over 2}\sum _{\langle i,j \rangle} \big[r_{i,j}-(R_i(\sigma_i)+R_j(\sigma_j))\big]^2, 
\label{eq_H}
\end{equation} 
where  $r_{i,j}$ is the distance between the centers of the $i$th and $j$th molecules. 
Each molecule takes the low spin (LS) state ($\sigma_i=-1$) or the high spin (HS) state ($\sigma_i=1$) and its radius $R$ depends on the state. 
 The LS molecule has a smaller radius than the HS molecule; $R_{\rm L}<R_{\rm H}$ where $R_{\rm L}$ ($R_{\rm H}$) is the radius of the LS (HS) molecule.
When the molecules contact each other, the energy of ${\cal H}_{\rm nn}$ is minimum. 
In Fig.~\ref{Fig_conf_T0}, the configurations of the minimum energy are depicted. 

To avoid a global deformation to rhombic shape and also to maintain the square lattice, a small perturbation, such as the following next nearest neighbor interaction, is necessary. 
\begin{equation}
{\cal H}_{\rm nnn}={k_2 \over 2}\sum _{\langle\langle i,k \rangle\rangle} \big[r_{i,k}-\sqrt{2}(R_i(\sigma_i)+R_k(\sigma_k))\big]^2. 
\label{eq_Hnnn}
\end{equation}
Then, the system of the elastic interaction has the form 
\begin{equation}
{\cal H}_{\rm Elastic}={\cal H}_{\rm nn}+{\cal H}_{\rm nnn}. 
\label{ela}
\end{equation}
We set $k_2=k_1/10$ in this study.  

Now, as the short-range interaction,
we introduce the following nearest-neighbor Ising interaction 
\begin{equation}
{\cal H}_{\rm IS}=-J\sum_{\langle i,j \rangle}\sigma_i \sigma_j.
\label{H_J}
\end{equation}

Taking into account the energy difference $D$ and 
the ratio $g$ of the degeneracies between the HS and LS states,\cite{Bousseksou2,Kamel2}
we also add the following term (see also Sec.~\ref{sec_Ising-like}), 
\begin{equation}
{\cal H}_{\rm eff}=\frac{1}{2}\left(D-{k_{\rm B}T}\ln g\right)\sum_i \sigma_i. 
\label{H_eff}
\end{equation}
Thus, the Hamiltonian of the system is given by
\begin{equation}
{\cal H}_{\rm tot}={\cal H}_{\rm Elastic}+{\cal H}_{\rm IS}+{\cal H}_{\rm eff}. 
\label{H_tot}
\end{equation}
In the present study, 
we focus on the critical phenomena. Thus, we need to investigate the system along the coexistence line of the model.  
It is approximately realized if we set $D-{k_{\rm B}T}\ln g=0$, i.e., $  {\cal H}_{\rm eff}=0$ (equivalent to $D=0$ and $g=1$). We give an analysis about the coexistence in Appendix~\ref{appendix_angle}.  
We set the parameter values as $R_{\rm H}=1.1$, $R_{\rm L}=1$, $k_1=40$ and $k_2=4$. 

Here we apply a Monte Carlo (MC) method with $NPT$ ensemble\cite{Konishi}, where the pressure is set $P=0$, for the square lattice (2D) with periodic boundary conditions. 
In the MC method, we choose a molecule at site $i$, and update the spin state 
$\sigma_i$ and the position of the molecule ($x_i,y_i$). 
Then we update the volume of the total system under $P=0$. 
One Monte Carlo step (MCS) is defined as $L \times L$ times of these procedures, where $L$ denotes the linear dimension (number of sites) of the system.

\subsection{Ground state configuration}

As mentioned above, some additional interaction
is necessary to maintain the square lattice. 
We here adopted the form (Eq. (\ref{eq_Hnnn}) ).
Other types of interactions can also be used to maintain the shape (see Appendix~\ref{appendix_angle}). 

Within ${\cal H}_{\rm nn}$, the ferromagnetic-like configuration (Fig.~\ref{Fig_conf_T0}(a) and (b)) and the antiferromagnetic-like configuration (Fig.~\ref{Fig_conf_T0}(c)) are degenerate.
However, ${\cal H}_{\rm nnn}$ resolves this degeneracy. 
The total energy of the ferromagnetic-like state per molecule 
is $E^{\rm F}_{\rm tot}/L^2=-Jz/2$, and that of the antiferromagnetic-like state is 
\begin{equation}
\frac{E^{\rm AF}_{\rm tot}}{L^2}=\frac{Jz}{2}+ \frac{k_2}{2}(\sqrt{2} (R_H-R_L))^2\frac{z}{2}, 
\end{equation}
where $z$ is the coordination number ($z=4$).
The energy difference between the two states is 
\begin{equation}
\frac{\Delta E}{L^2}=\frac{E^{\rm AF}_{\rm tot}-E^{\rm F}_{\rm tot}}{L^2}=Jz+\frac{k_2(R_H-R_L)^2 z}{2}. 
\end{equation}
Therefore, for $J=0$, $\Delta E>0$ and the ferromagnetic-like state is favored.  
Substituting $R_H=1.1$, $R_L=1.0$, $z=4$, and $k_2=4$, $\Delta E/L^2=4(J+0.02)$. 
Thus, at $J=J_0=-0.02$, the ground state changes between the ferromagnetic-like ($J>J_0$) and antiferromagnetic-like ($J<J_0$) states. 
Hereafter, we define $J_0$ as the origin of $J$.

\section{Critical properties of two order parameters}
\label{secthir}

We study the dependence of the critical properties of the model on the short-range interaction $J$. 
In the present model,
the magnetization ($m$) and staggered magnetization ($m_{\rm sg}$) are the essential order parameters. 
The definitions of $m$ and $m_{\rm sg}$ are given by  
\begin{equation}
m=\frac{\sum_i \sigma_i}{L^2} 
\end{equation}
and
\begin{equation}
m_{\rm sg}=\frac{\sum_i (-1)^{i_x+i_y} \sigma_i}{L^2}, 
\end{equation}
where ($i_x$, $i_y$) is the integer coordinate of the $i$th molecule which numbers the 2D lattice. It should be noted that $m$ ($m_{\rm sg}$) is not real (staggered) magnetization but a kind of pseudo (staggered) magnetization to show ferromagnetic-like (antiferromagnetic-like) order.  

In order to study the critical phenomena, we analyze Binder cumulants for both order parameters. 
Binder cumulants for ferromagnetic-like and antiferromagnetic-like orders are, respectively, defined as
\begin{equation}
U_4^{\rm F}(L)\equiv  1-{\langle m^4\rangle_L\over 3\langle m^2 \rangle^2_L}
\;\;\; {\rm and } \;\;\;
U_4^{\rm AF}(L)\equiv 1-{\langle m_{\rm sg}^4\rangle_L\over 3\langle m_{\rm sg}^2 \rangle^2_L}. 
\label{Binder cum.}
\end{equation} 
At the critical temperature, the Binder cumulants for different values of $L$ cross, and the value at the point depends on the type of phase transition.

We also investigate the correlation function of the spin state in the vicinity of the critical point to catch the difference of the ordering between ferromagnetic-like and antiferromagnetic-like phases. 
The definition of the correlation function is given by
\begin{equation}
C(i,j)= \langle \sigma_{(l_x,l_y)} \sigma_{(l_x+i,l_y+j)}  \rangle.   
\end{equation}
Here ${(l_x,l_y)}$ denote the position of the $l$-th spin, and $i$ and $j$ are taken in the range $0 \leq i, j \leq L/2-1$. 

\begin{figure}[th]
  \begin{center}
     \includegraphics[width=50mm]{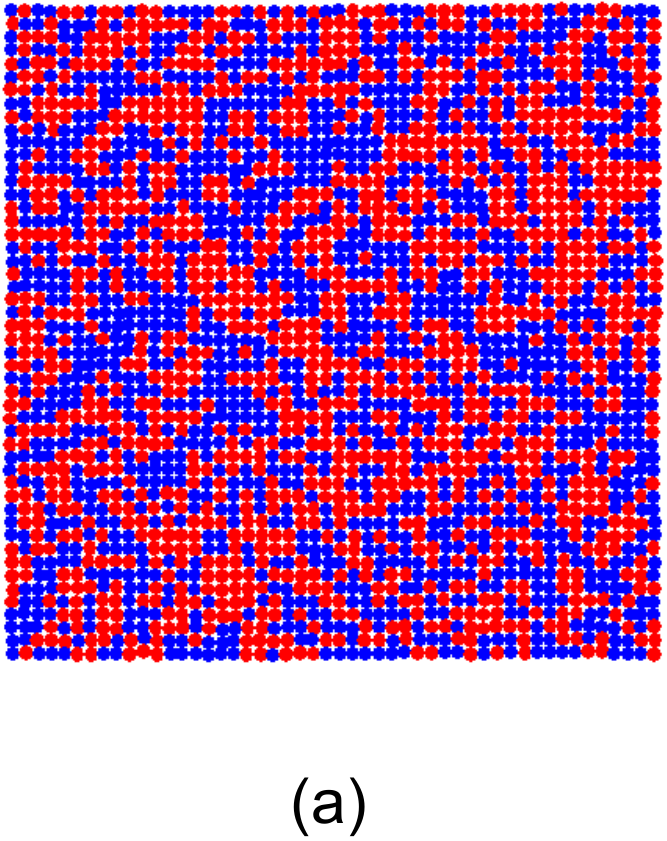}
     \hspace{3cm}
     \includegraphics[width=70mm]{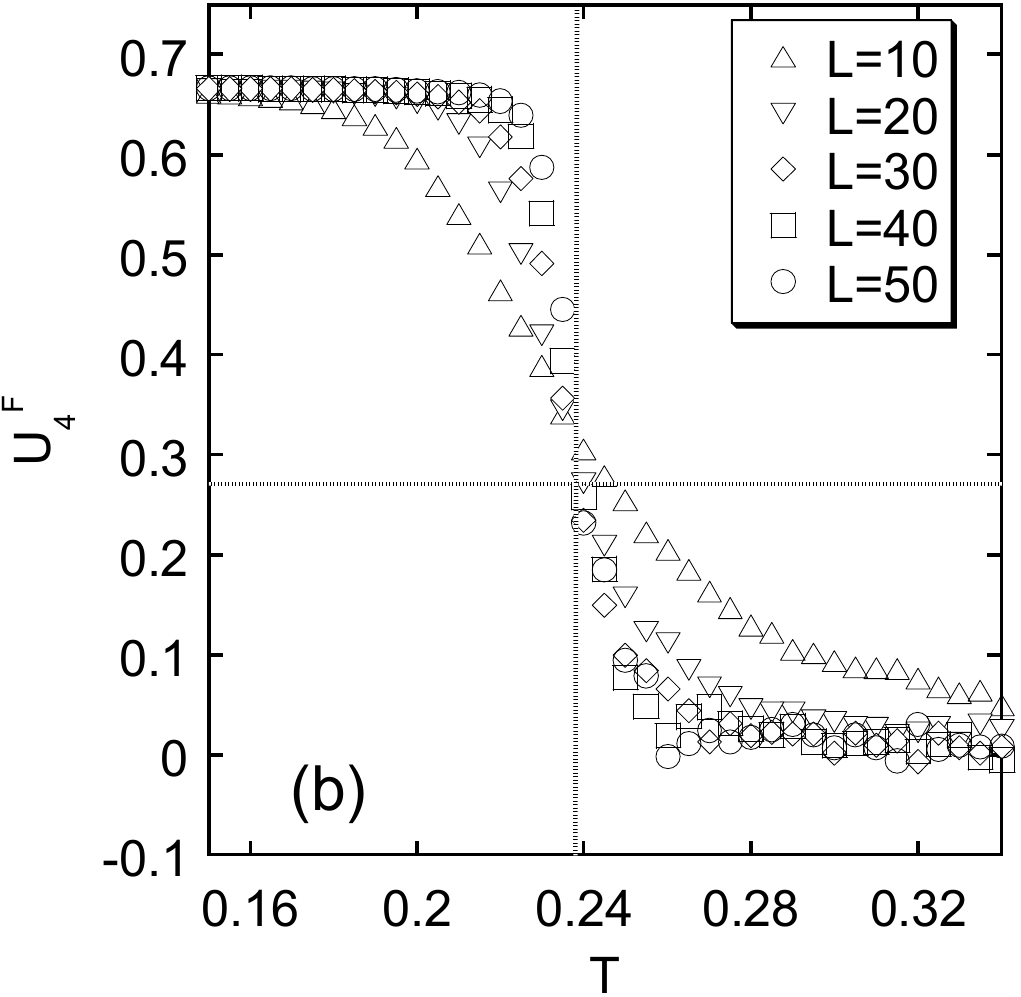}
  \end{center}
\caption{(color online) (a) A snapshot of the configuration at $T=0.24$ (near $T_{\rm c}=0.238$) for $J=0.01$ and $L=50$. A uniform configuration is seen. 
(b)  Temperature dependence of $U^{\rm F}_4$  for various system sizes $L$ when $J=0.01$. The horizontal dotted line ($U^{\rm F}_4=0.27$) denotes the fixed-point value of the Binder cumulant for the mean-field universality class. 
}
\label{Fig_ferromag}
\end{figure}

\subsection{Ferromagnetic-like phase transition}
\label{secfour}

First we study the case of the ferromagnetic-like phase transition.
Here we consumed 1000,000-2000,000 MCS for the initial equilibration and the following 1000,000-6000,000 MCS at each temperature to obtain physical quantities.

Figure~\ref{Fig_ferromag} (a) shows a typical configuration near the critical temperature $T_{\rm c}$ when $J=0.01$. The red (blue) solid circles denote HS (LS) molecules. The characteristic features are the same as we studied for $J=0$,\cite{Miya} and no clustering occurs. 
In Fig.~\ref{Fig_ferromag} (b), the Binder plot, $U^{\rm F}_4(L)$ (Eq.~\ref{Binder cum.}), is given  for several system sizes $L$.
We estimate the critical temperature of the ferromagnetic-like phase transition 
from the crossing point: $T_{\rm c}=0.238$. 
The value of $U^{\rm F}_4(L)$ at the point agrees well with that of the mean-filed model: 
 $U^{\rm F}_4(L)=1-\Gamma^4(1/4)/24\pi^2 =0.27...$~\cite{Brezin,Luijten} 
Thus we conclude that 
the elastic model with $J>J_0$, which shows the ferromagnetic-like transition belongs to the mean-filed universality class. 
This is consistent with a previous study.\cite{Nakada2} 
(In Sec.~\ref{sec_phase_diagram}, we discuss in detail the situation when $J$ is closer to $J_0$.)

\subsection{Antiferromagnetic-like phase transition}

\begin{figure}[H]
  \begin{center}
     \includegraphics[width=130mm]{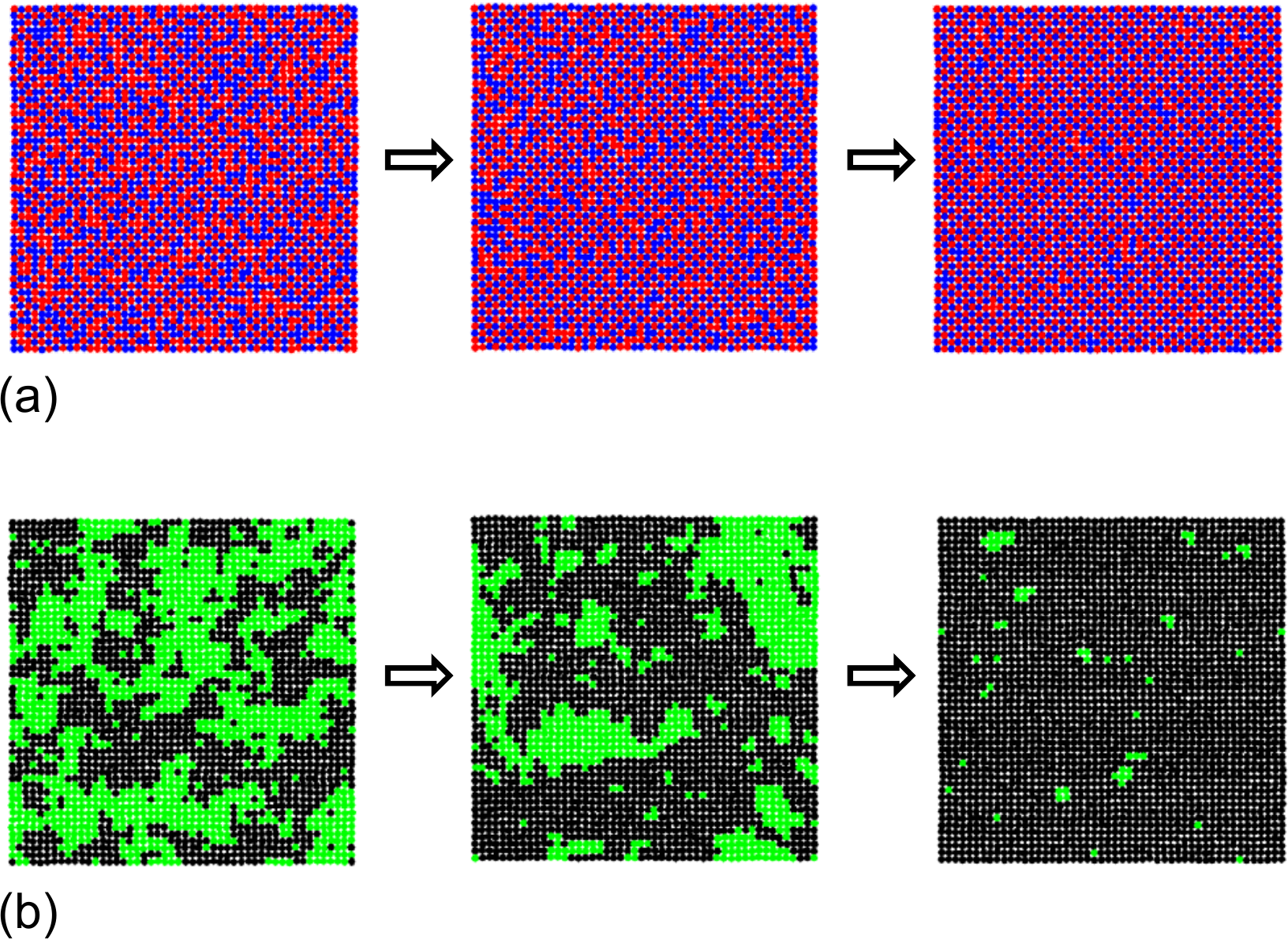}
  \end{center}
%\vspace{0.5cm}
  \begin{center}
     \includegraphics[width=70mm]{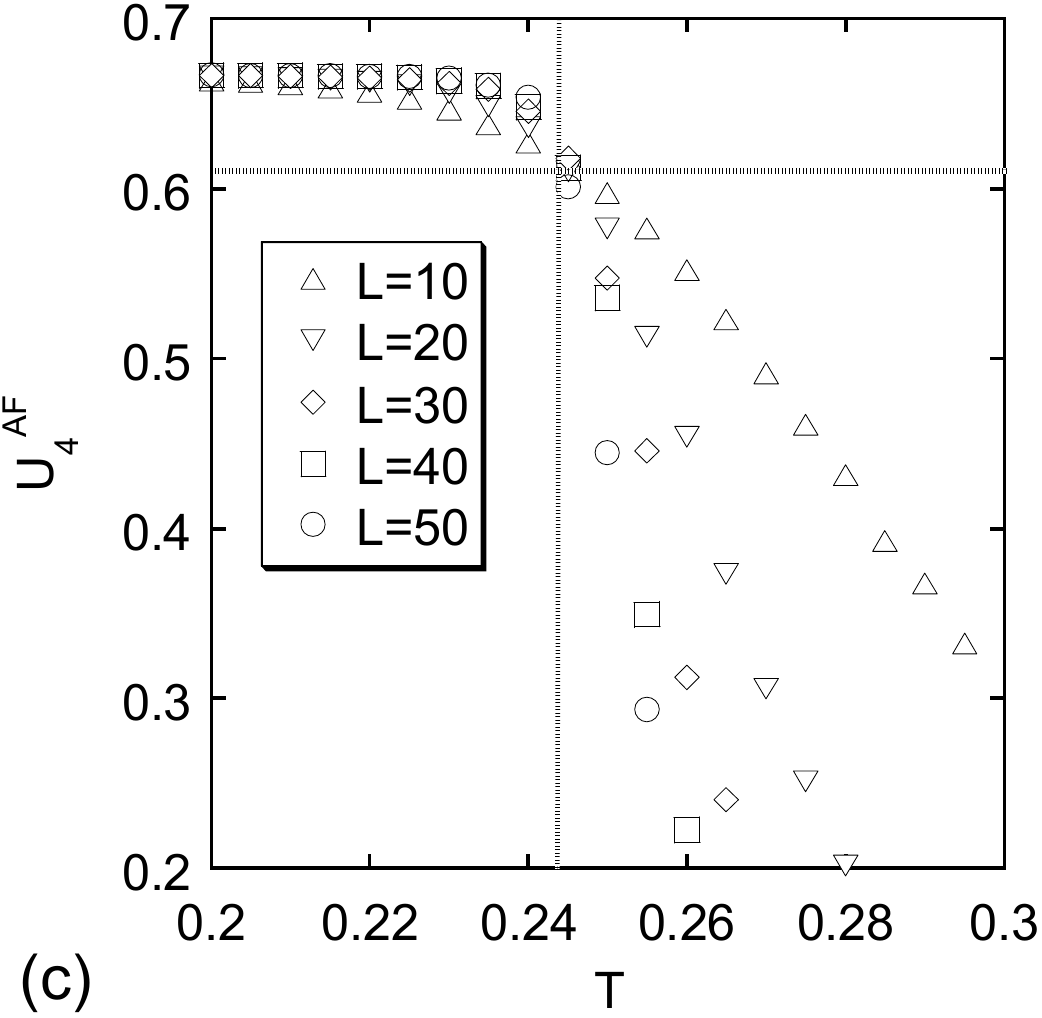}
  \end{center}
\caption{(color online) (a) Snapshots of the configuration for $J=-0.1$ and $L=50$.  Small clusters of the antiferromagnetic-like phase appear at the left ($T=0.29$),  
large clusters appear at the middle ($T=0.25$) near the critical point $T_{\rm c}=0.243$, and antiferromagnetic-like ordered phase is realized at the right ($T=0.2$). 
(b) Snapshots of the masked configuration of (a). 
(c) Temperature dependence of $U^{\rm AF}_4$  for various system sizes $L$ when $J=-0.1$. The horizontal dotted line ($U^{\rm AF}_4=0.61$) denotes the fixed-point value of the Binder cumulant for the Ising universality class. 
}
\label{Fig_antiferro}
\end{figure}

Next we study the case of the antiferromagnetic-like phase transition.
For $J < -0.02$ we expect that an antiferromagnetic-like transition takes place. 
Here we study the case of $J=-0.1$.
Snapshots of the configuration for $L=50$ are given in Fig.~\ref{Fig_antiferro} (a). 
In contrast to the case of ferromagnetic-like order, the clusters consist of alternating LS and HS configurations. 
To find the clusters more easily, we also plot masked configurations of Fig.~\ref{Fig_antiferro}  (a) in Fig.~\ref{Fig_antiferro} (b). Here the masked configuration is given by $\sigma'_i={(-1)^{i_x+i_y} \times \sigma_i}$ and black (green) circles denote $\sigma'_i=1$ $(-1)$. 
There, small antiferromagnetic-like domains are observed at $T=0.29$ (left) and large antiferromagnetic-like domains at the middle ($T=0.25$ which is close to the critical temperature $T_{\rm c}=0.243$). An ordered antiferromagnetic-like phase is observed at $T=0.2$ (right).

The Binder plot is given in Fig.~\ref{Fig_antiferro} (c) for several system sizes $L$. We find $T_{\rm c}=0.243$ from the crossing point, 
and the value of $U_4^{\rm AF}(L)$ at the point agrees well with that of the short-range Ising model: $U^{\rm F}_4(L)=0.61...$~\cite{IsingCum} 
Unlike the case of the ferromagnetic-like transition, the elastic model with $J<J_0$ showing the antiferromagnetic-like transition belongs to the short-range Ising universality class.

%%%---------------------------------------------------------------------

\subsection{Correlation function}
\label{sec_corr}

We investigate the correlation function $C(i,j)/C(1,1)$, in which the value $C(1,1)$ is chosen as the reference value, and study the temperature dependence of the ordering patterns near $T_{\rm c}$ in both cases of the ferromagnetic-like and antiferromagnetic-like transitions.

\begin{figure}[H]
  \begin{center}
     \includegraphics[width=120mm]{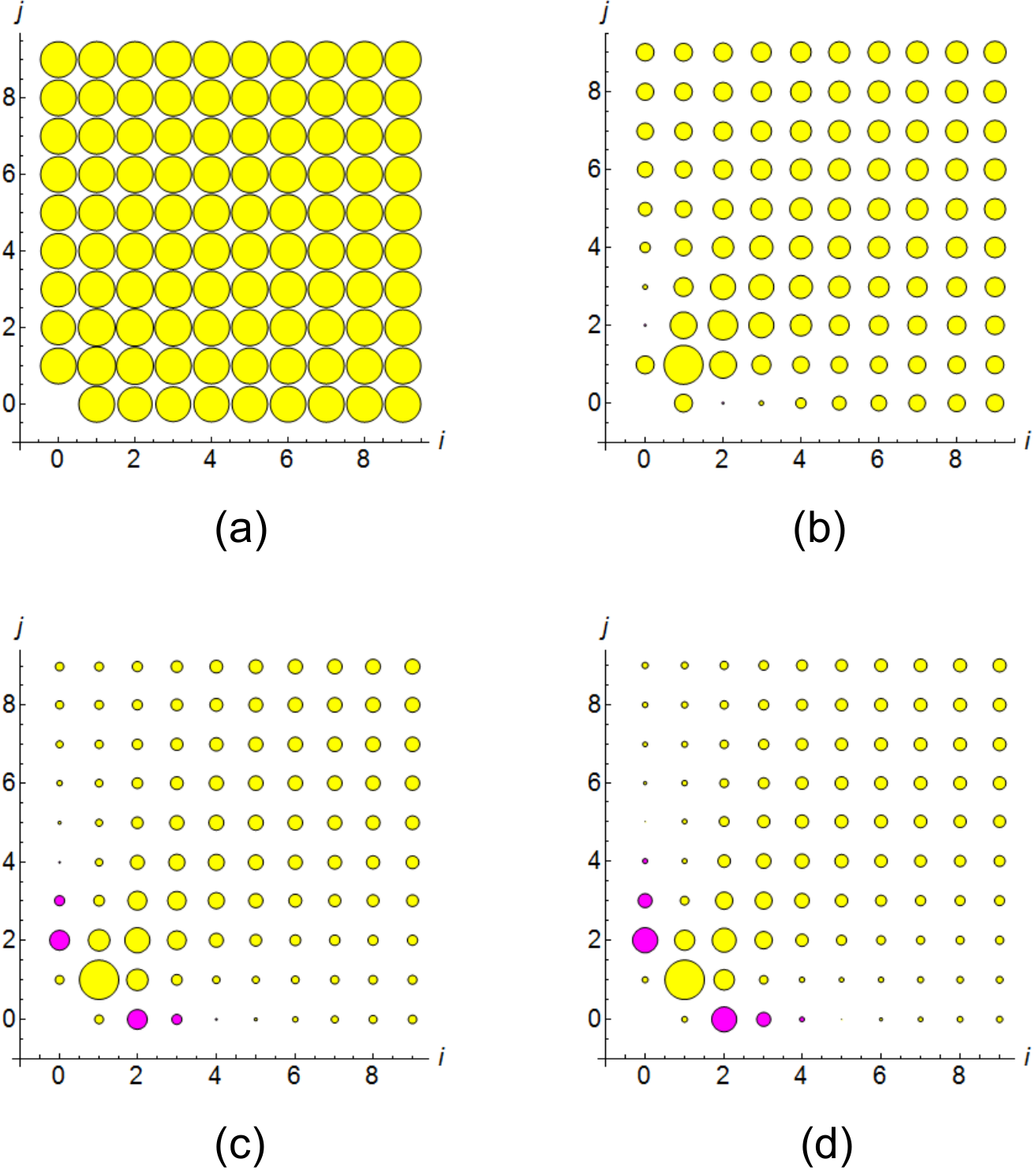}
  \end{center}
\caption{(color online)  Profiles of the correlation function of the system with $J=0.01$ and $L=20$ around $T_{\rm c}=0.238$. (a) $T=0.200$, (b) $T=0.240$, (c) $T=0.260$ and (d) $T=0.270$. The diameter of each disk corresponds to the value of $C(i,j)/C(1,1)$. Yellow (Magenta) means the plus (minus) sign of the correlation function.  }
\label{corr_f}
\end{figure}

\begin{figure}[H]
  \begin{center}
     \includegraphics[width=120mm]{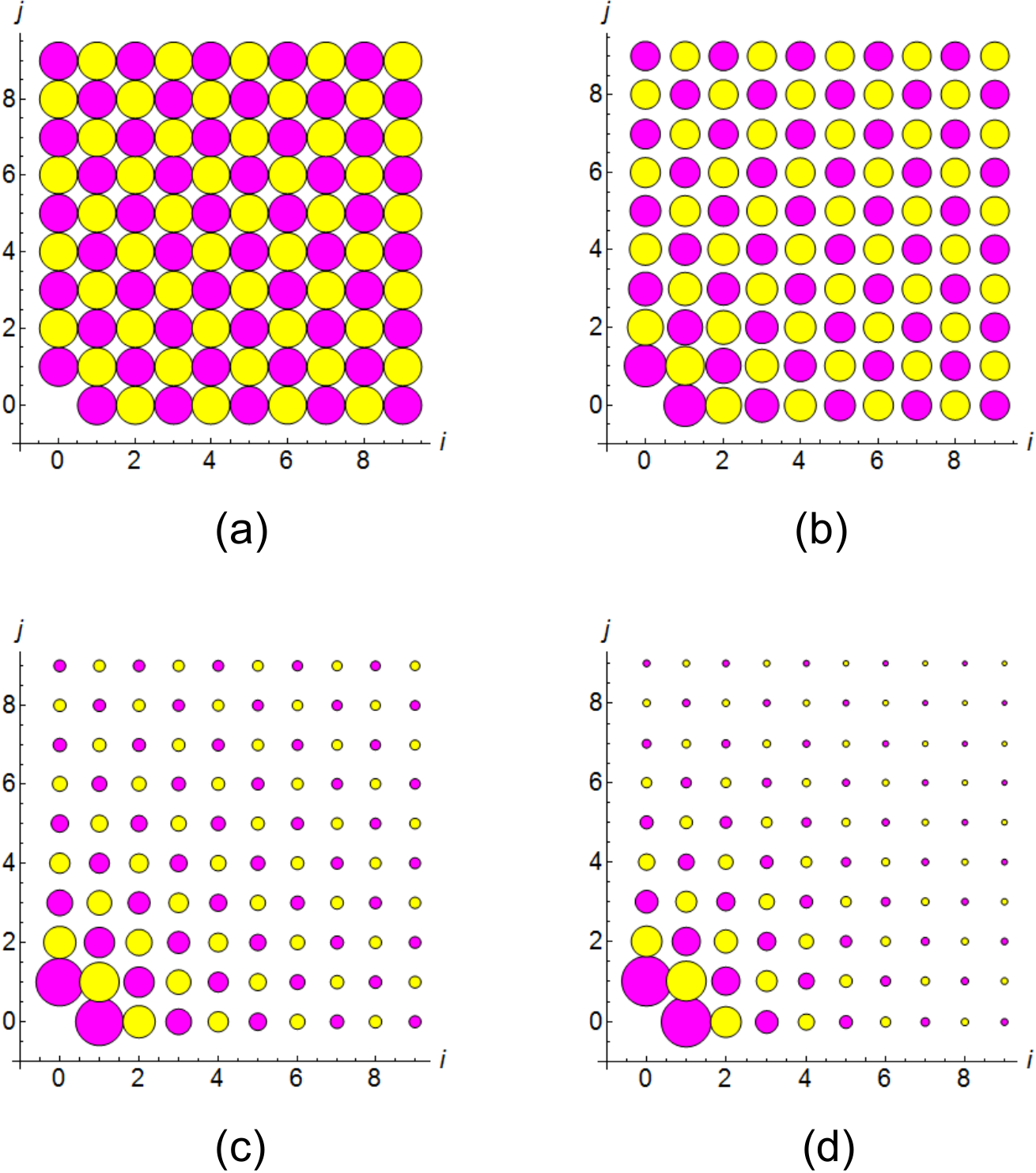}
  \end{center}
\caption{(color online)  Profiles of the correlation function $C(i,j)/C(1,1)$ of the system with $J=-0.1$ and $L=20$ around $T_{\rm c}=0.243$. (a) $T=0.205$, (b) $T=0.245$, (c) $T=0.265$ and (d) $T=0.275$. 
The diameter of each disk corresponds to the value of $C(i,j)/C(1,1)$. 
Yellow (Magenta) means the plus (minus) sign of the correlation function. 
}
\label{corr_af}
\end{figure}

Figures~\ref{corr_f} and \ref{corr_af} illustrate profiles of the correlation function when $J=0.01$ and $J=-0.1$, respectively. 
The former shows the change of the correlation function in ferromagnetic-like transition, while the latter shows it in antiferromagnetic-like transition. 
In both transitions, 
(a) depicts a configuration below $T_{\rm c}$, 
(b) a configuration close to $T_{\rm c}$, 
(c) and (d) configurations above $T_{\rm c}$.  

The configurations (c) and (d) in Fig.~\ref{corr_f} demonstrate that 
the correlation is still large at long distances in the ferromagnetic-like case, which is a characteristic of long-range interaction systems.~\cite{Miya} 
In contrast to this feature, the correlation decreases rapidly at long distances in the antiferromagnetic-like case as shown in Figs~\ref{corr_af} (c) and (d). 
This is characteristic of short-range interaction models like the Ising model. 
This analysis of the correlation functions is consistent with the analysis of the Binder cumulants, i.e., the universality class is different in the two cases. The difference between the mean-field and Ising universality classes is confirmed by the correlation function. 

What causes the difference of the universality class between ferromagnetic-like and antiferromagnetic-like ordering? 
If we consider a configuration in which the LS phase and HS phase coexist with a domain wall, the interface between the two phases causes an energy cost of the order of $O(L^2)$, which is the same mechanism as $L^2$-dependence of the activation energy in the macroscopic nucleation of elastic interaction systems in 2D.\cite{Macro_nuc} (In the case of 3D, the energy cost is of order $O(L^3)$.)  
In this case, distortions of the lattice are proportional to the size of the cluster. 
Therefore, ferromagnetic-like large clustering is impossible because of the huge energy cost. As a result, the uniform configuration during the transition is universal in ferromagnetic-like ordering. 

The situation is different in the antiferromagnetic-like case. In appearance of the antiferromagnetic-like phase, the symmetry is broken between one configuration of LS, HS, LS, HS.. and the other configuration HS, LS, HS, LS..... If we consider a joint system of these two antiferromagnetic-like phases, the interface energy is of the order of $O(L)$ because these two antiferromagnetic-like phases have the same unit area in 2D (In the case 3D, the energy cost is of $O(L^2)$ because of the same unit volume).  
Thus, the size difference of the LS and HS molecules essentially does not cause energy cost. Therefore, antiferromagnetic-like ordering accompanied by clustering is possible as is the case of usual phase transitions in short-range interaction models. 


\clearpage

\begin{figure}[t]
  \begin{center}
     \includegraphics[width=90mm]{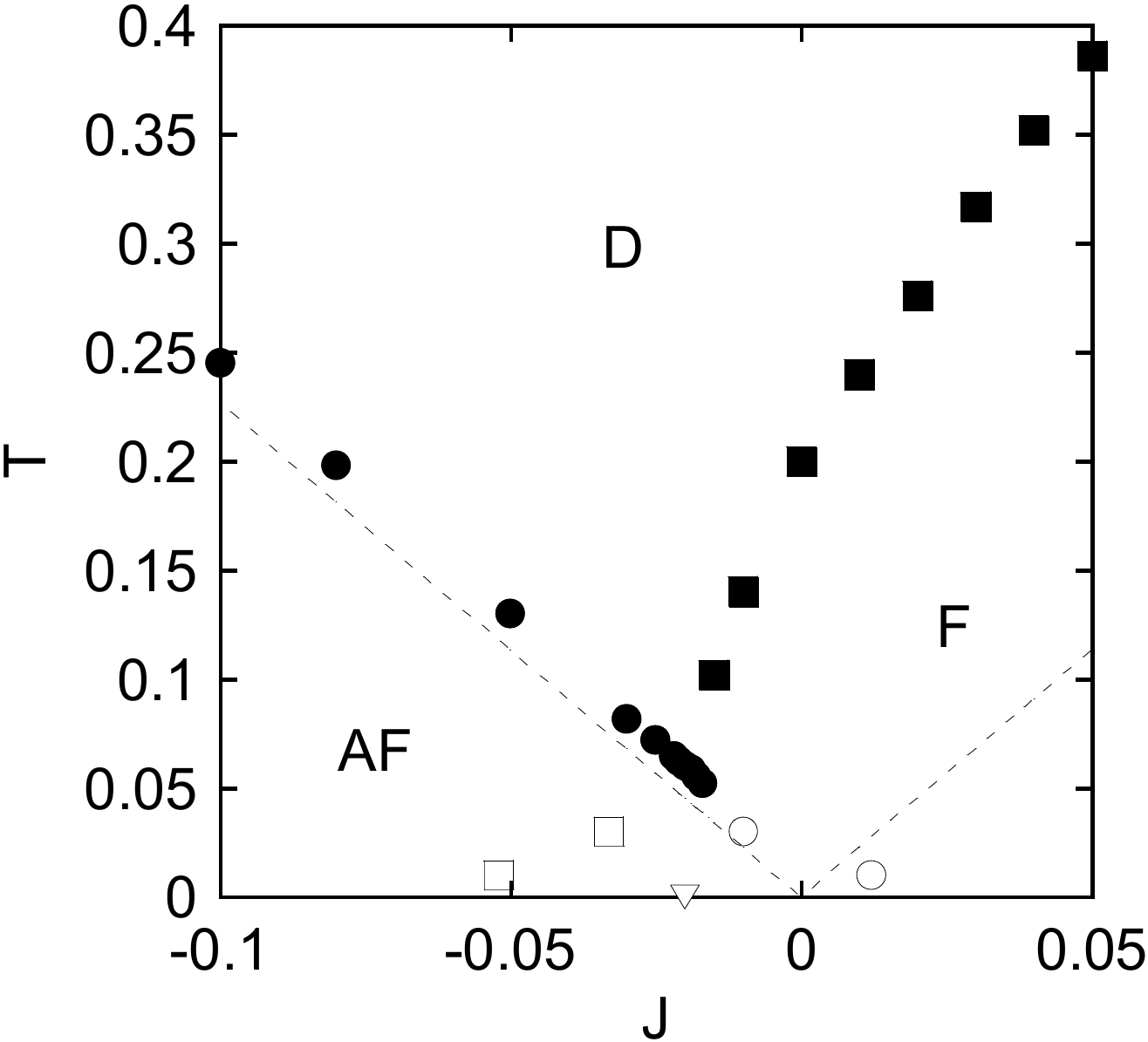}
  \end{center}
\caption{Phase diagram in terms of the short-range interaction $J$ vs temperature $T$. F denotes the region of the ferromagnetic-like phase, AF the region of the antiferromagnetic-like phase, and D the region of the disordered phase. 
The point $J=J_0=-0.02$ at $T=0$  (downward-triangle) is the critical point between the ferromagnetic-like and antiferromagnetic-like phases at $T=0$. 
Solid circles and squares denote the critical points for the antiferromagnetic-like and ferromagnetic-like transitions, respectively. 
The dashed line shows the critical temperature given by only the Ising interaction $J$. 
Open squares and circles denote the end points of the metastable ferromagnetic-like and antiferromagnetic-like phases (see Fig.~\ref{Fig_metastability}), respectively. 
}
\label{Fig_phase_diagram}
\end{figure}

\section{Phase diagram}
\label{sec_phase_diagram}

We depict the phase diagram $J$ vs. $T$ in Fig.~\ref{Fig_phase_diagram}. 
Solid circles and solid squares denote critical temperatures ($T_{\rm c}$) of antiferromagnetic-like and ferromagnetic-like phases, respectively. 
As pointed out in sec.~\ref{sec2}, the original point between the antiferromagnetic-like and ferromagnetic-like phases in the ground state is $J=-0.02$. 
It is worth nothing that the antiferromagnetic-like phase transition occurs at the point $J=J_0(=-0.02) $ although the ferromagnetic-like and antiferromagnetic-like phases are degenerate at $T=0$ for $J_0$. 
We find that the antiferromagnetic-like phase transition occurs even for $J \le -0.017$, larger than $J=J_0$. It is considered as one of the reasons that the average density of the disordered phase is closer to that of the antiferromagnetic-like phase than that of the ferromagnetic-like phase, and thus the generation of antiferromagnetic-like clusters is easier to realize than the generation of the ferromagnetic-like phase which is accompanied by global volume change.

Considering that the critical temperature of the pure Ising model\cite{2D_Ising} ( Eq.(\ref{H_J})) (for both ferro and antiferromagnetic cases) is given by
\beq 
T_{\rm c}=\frac{2}{\ln (1+ \sqrt{2})}|J| \simeq 2.27|J|, 
\eeq
it is found that the critical temperatures of the antiferromagnetic-like transition, shown by solid circles in Fig.~\ref{Fig_phase_diagram}, are a little bit larger than the temperatures $T=2.27|J|$ which is shown by the dashed lines in Fig.~\ref{Fig_phase_diagram}. 
The critical temperature of the antiferromagnetic-like transition can be approximated by 
\beq 
T_{\rm c}\simeq 2.27 (|J|+C),  
\eeq
where $C \simeq 0.01$. 
Here it should be noted that $J_0$ is just the dividing point between the ferromagnetic-like and antiferromagnetic-like phases at $T=0$, and the critical temperature $T_{\rm c}$ is a function of $J$ (not $J-J_0$) because the interface energy is a function of $J$. 

We consider the origin of the shift $C$ of the critical temperature. We may attribute it to the cost of the elastic interaction at the interface. 
At the interface in $x$ or $y$ direction  between two phases (LS, HS, LS, HS... and HS, LS, HS, LS, ...),  LS-LS molecular pair of the nearest neighbors (call LS-LS bond) and HS-HS molecular pair of the nearest neighbors (HS-HS bond) align alternately. 
The ideal intermolecular distance $r_{i,j}$ which gives the minimum energy is different for LS-LS and HS-HS pairs. The surface energy due to the elastic interaction is calculated in a simple approximation as follows. 
HS-HS pair favors the distance $r_{i,j}=2 R_{\rm H}$ but LS-LS pair favors $r_{i,j}=2 R_{\rm L}$. Assuming that $r_{i,j}=R_{\rm L}+R_{\rm H}$ is realized as a result of compromise, the energy costs per pair is $\Delta E \simeq  {k_1 \over 2}[r_{i,j}-(R_i(\sigma_i)+R_j(\sigma_j))]^2 \simeq  {k_1 \over 2}[(R_{\rm L}+R_{\rm H})-2 R_{\rm L}]^2={k_1 \over 2}[(R_{\rm L}+R_{\rm H})-2 R_{\rm H}]^2={k_1 \over 2}[R_{\rm H}-R_{\rm L}]^2$=0.2, which leads to $C \simeq 0.1$. However, the true value of $C$ is much smaller and the relaxation of the configuration would  be necessary. 

Then we estimate the elastic interface energy as follows. 
We calculated the difference of the elastic energies (Eq.~\ref{ela}) between the system with no interface (complete antiferromagnetic-like phase) and that with an interface. For $L=20$ and $L=40$, we obtained relaxed configurations for both systems at $T=0$ and estimated the elastic energies. 
Dividing the difference of the elastic energies by the number of LS-LS and HS-HS bonds on the interface, we have $\delta E \simeq 0.023$ (per interface bond) in the case of the interface in $x$ (or $y$) direction for both $L=20$ and $L=40$. 
We also estimated $\delta E$ in the case of the interface in the diagonal direction, where interface line consists of LS-LS bonds or HS-HS bonds, and found $\delta E$ much smaller.  Thus $C$ is considered the order of 0.01. 
The energy cost due to interfaces is released to elastic interactions around the interfaces.  The elastic interactions enhance antiferromagnetic-like ordering weakly. 

In contrast, in the ferromagnetic-like transition, the effective long-range interaction significantly enhances the transition temperature and the values of $T_{\rm c}$ are much larger than those of the antiferromagnetic-like transition. 
Unlike the antiferromagnetic-like transition, the critical temperature increases much more steeply than 2.27$J$ with the increase of the value of $J$. 
This indicates that the elastic interaction and the short-range interaction $J$ enhance ferromagnetic-like ordering synergetically with non-linear dependence.

\begin{figure}[t]
  \begin{center}
     \includegraphics[width=130mm]{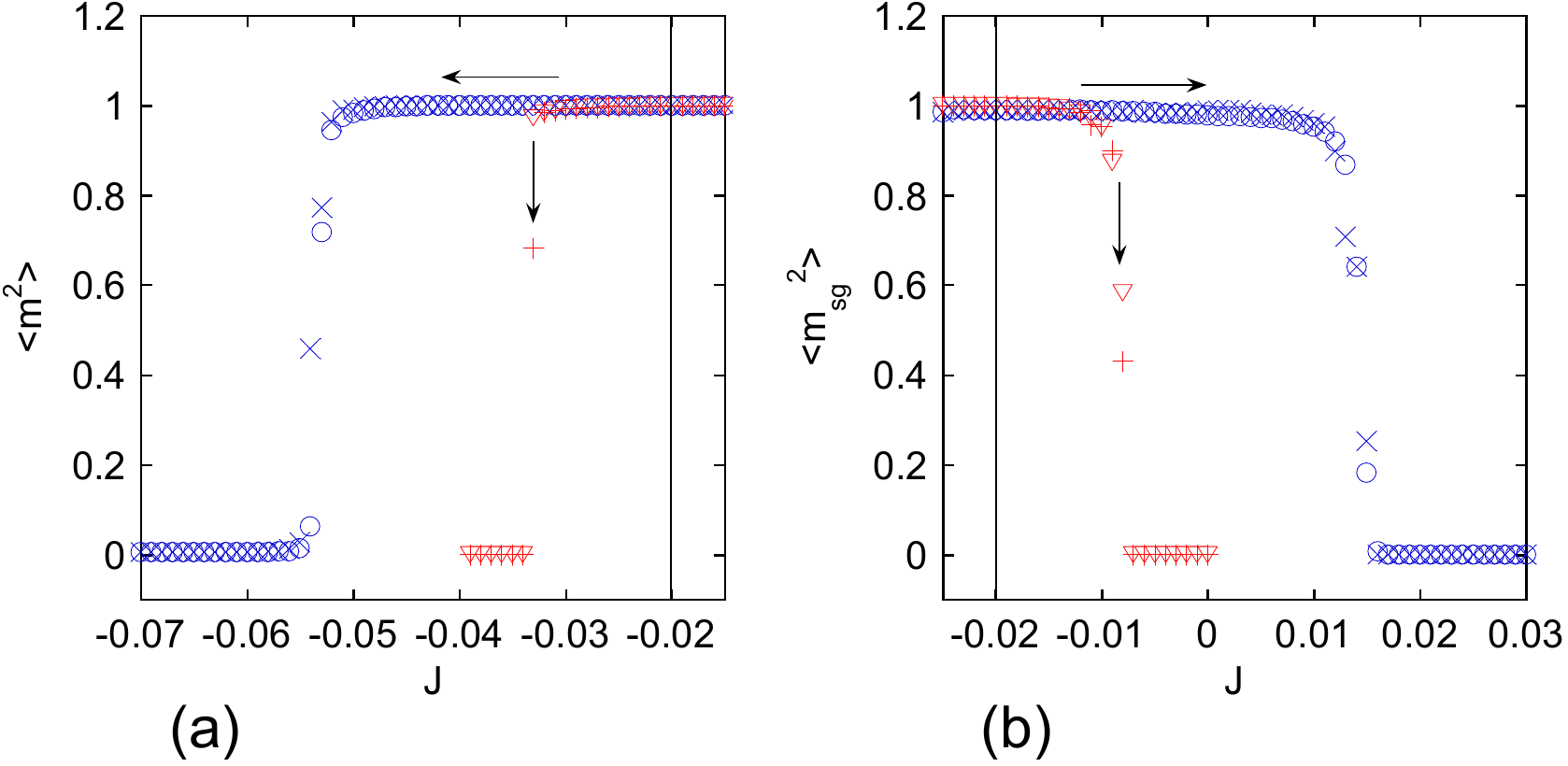}
  \end{center}
\caption{(color online) (a) Metastability of the ferromagnetic-like phase for the values of $J$, 
(b) Metastability of the antiferromagnetic-like phase for the values of $J$. 
 $+$ and $\times$ denote the order parameters at $T=0.03$ and $T=0.01$, respectively for $L=20$. 
 $\bigtriangledown$ and $\bigcirc$ denote the order parameters at $T=0.03$ and $T=0.01$, respectively for $L=40$. 
 }
\label{Fig_metastability}
\end{figure}

It is expected that around $J_0$ the ferromagnetic-like and antiferromagnetic-like orders are nearly degenerate which causes a metastable structure of the ordered states. 
We study the metastable regions by the analysis of the dynamics of the order parameters under a sweep of $J$ (Figs.~\ref{Fig_metastability} (a) and (b)). 

In Fig.~\ref{Fig_metastability} (a), we set ferromagnetic-like phase as the initial state at temperatures ($T=0.01, 0.03$), and decreased gradually the value of $J$ from $-0.01$ to $-0.07$, and observed the relaxation of $\langle m^2\rangle$.  We identified the point where 
$\langle m^2\rangle$ decreased large as the end-point of the ferromagnetic-like metastable region.
Here, in the process of decreasing of the value of $J$, we changed $J$ 
in steps of 0.001 and used 1000,000 MCS to equilibrate and the following 1000,000 MCS to measure $\langle m^2 \rangle$ for each $J$. 
We determined the boundary of the metastable region 
where $\langle m^2 \rangle$ becomes smaller than $0.9$. 
We show the data of average over five trials with the use of different random number sequences for $L=20$ and $L=40$. We do not find strong dependence on $L$ for the metastability and conclude that the border of metastable state is well defined. These points are plotted by open squares in Fig.~\ref{Fig_phase_diagram}. 

In Fig.~\ref{Fig_metastability} (b), we set antiferromagnetic-like phase as the initial state at temperatures ($T=0.01, 0.03$), and increased gradually the value of $J$ from $-0.03$ to $0.03$, and observed the relaxation of $\langle m_{\rm sg}^2\rangle$.  We identified the metastable antiferromagnetic-like phase in the same way for the ferromagnetic-like case. The estimated boundary points are plotted by open circles in Fig.~\ref{Fig_phase_diagram}. 
We find that metastability is realized and the region expands at lower temperatures.  

\section{Outline of Spin crossover transitions}
\label{sec_Ising-like}

To study the characteristics of 
two-step (HS $\leftrightarrow$ AF $\leftrightarrow$ LS) SC transitions on the basis of the analyses in the previous sections, we briefly summarize in this section the essential features of the types of SC transitions making use of 
the phase diagrams for the conventional (Ising-like) models. 
We will discuss the difference of the critical properties between the elastic model and Ising-like models in Sec.~\ref{summary}.

Ising-like models have been developed to describe various SC phenomena. The Ising model with multi-fold degeneracy was studied for single-step (LS $\leftrightarrow$ HS) spin-crossover transitions to catch the entropy induced transitions.~\cite{WP} 
The model is given by 
\begin{equation}
{\cal H}=-J_{\rm F} \sum_{\langle i,j \rangle}\sigma_i \sigma_j + 
\frac{D}{2} \sum_i \sigma_i, \;\;\;\sigma_i=\underbrace{-1,\cdot \cdot ,-1,}_{u}\underbrace{1,\cdot \cdot \cdot \cdot 1}_{r}.  
\label{HamWP}
\end{equation}
Here $J_{\rm F}>0$ denotes the interaction (not magnetic) between the nearest neighbor molecules and $D>0$ denotes the energy difference between the LS state ($\sigma_i=-1$) and the HS state ($\sigma_i=1$). The numbers of degenerate states $u$ and $r$ are associated with the LS and HS molecular states. 
The Hamiltonian (\ref{HamWP}) is equivalent to the following one in the partition function.~\cite{Kamel2} 
\begin{equation}
{\cal H}=-J_{\rm F} \sum_{\langle i,j \rangle}\sigma_i \sigma_j + 
\frac{1}{2}\left(D-{k_{\rm B}T}\ln g\right)\sum_i \sigma_i, \;\;\;\sigma_i=-1, \; 1, 
\label{Ising-like}
\end{equation}
where $T$ is temperature and $g \equiv \frac{r}{u}>0$. The second term is ${\cal H}_{\rm eff}$ in Eq.~(\ref{H_eff}). 
This is the Ising model with an effective field $h(T)=-\frac{1}{2}\left(D-{k_{\rm B}T}\ln g\right)$. 

\begin{figure}[h]
  \begin{center}
     \includegraphics[width=53mm]{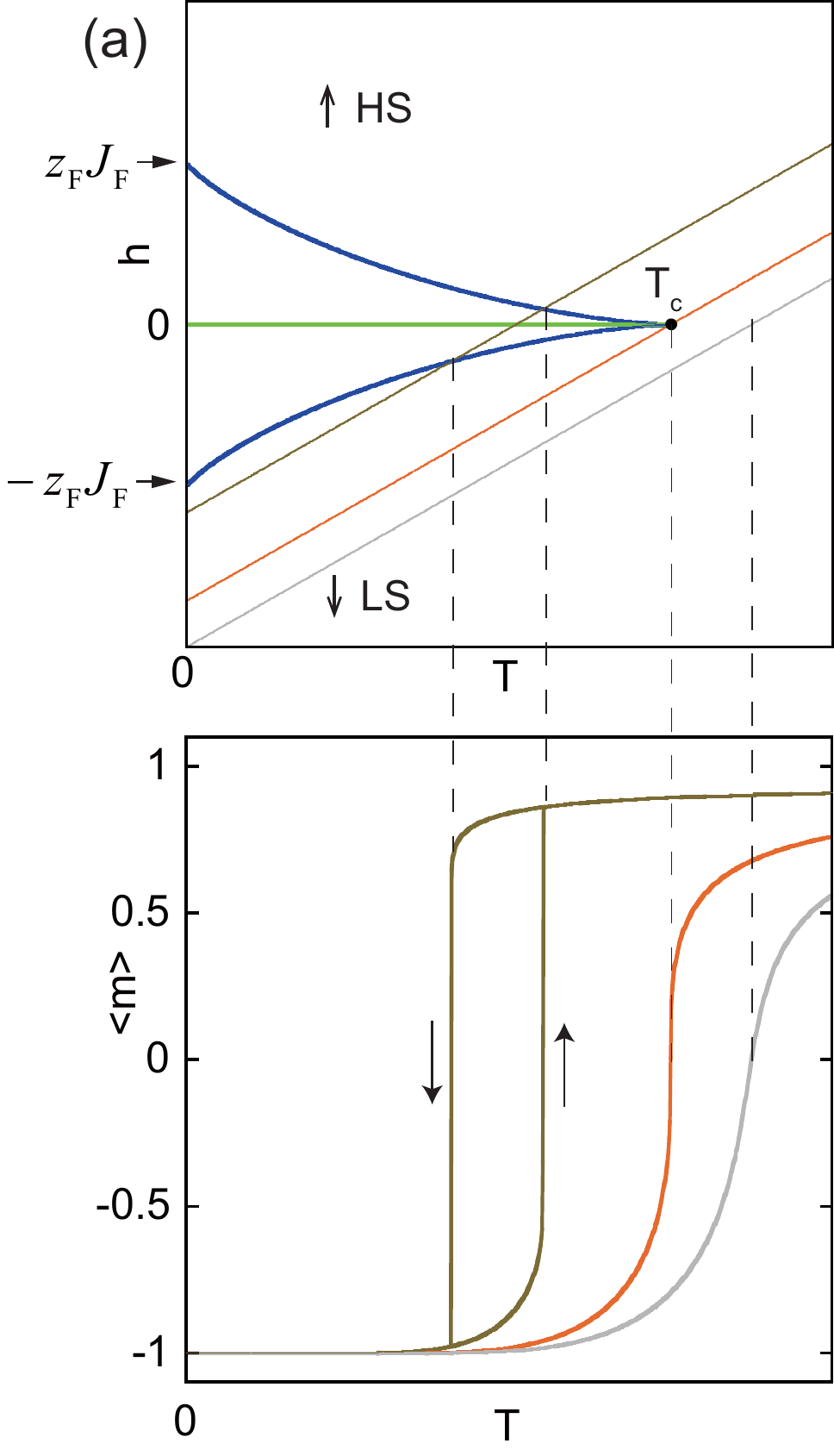}
\hspace{0.5cm}
     \includegraphics[width=56mm]{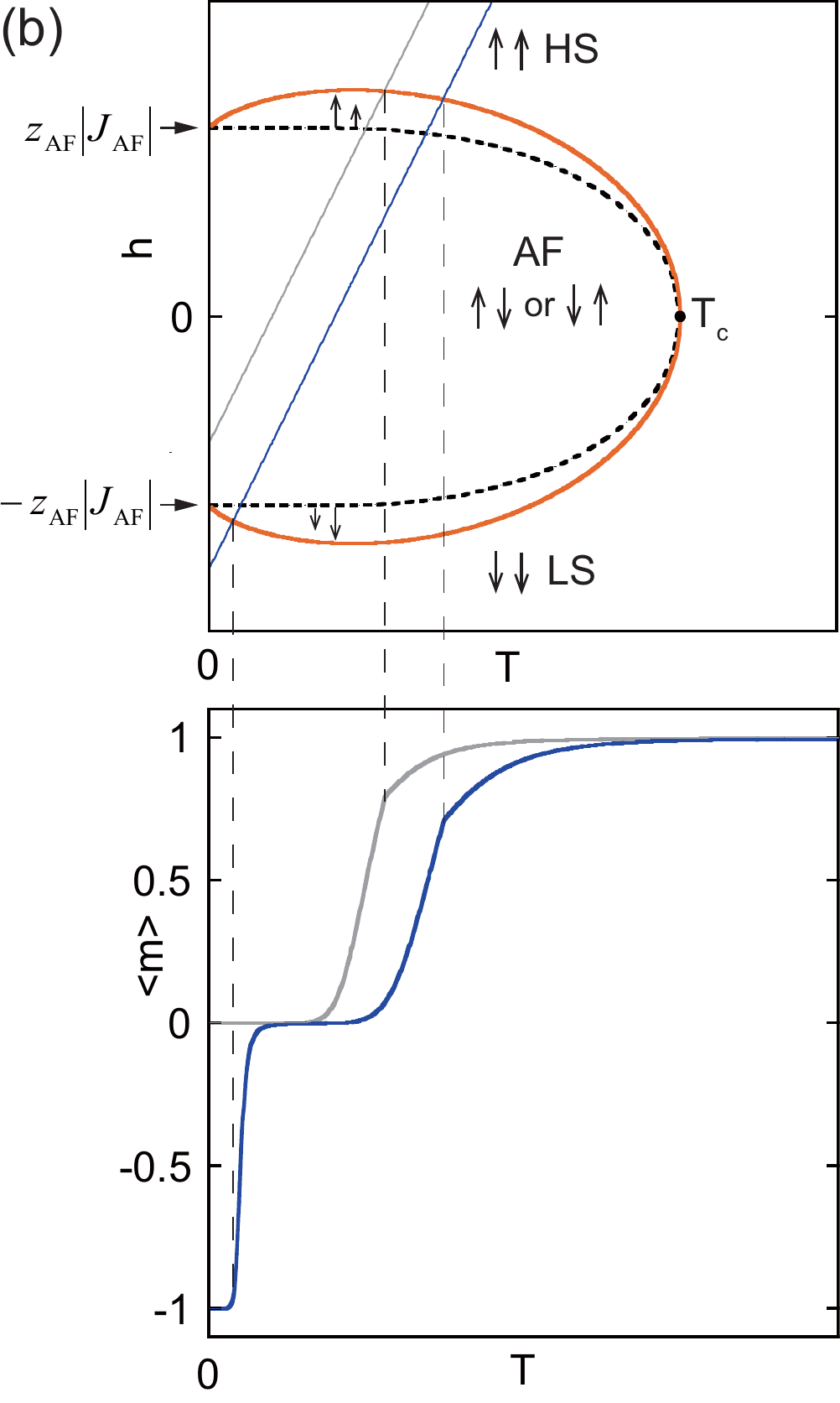}
  \end{center}
\caption{(color online)  (a) Phase diagram of the Ising model for single-step SC transitions (upper panel) by the mean-field theory and three typical cases of temperature dependence of $\langle m \rangle$ (lower panel). Spinodal curves are shown by blue lines and coexistence line by a green line. The critical temperature $T_{\rm c}$ is given by $T_{\rm c}=z_{\rm F} J_{\rm F}$ in the mean-field theory. Three cases of first-order, second-order transition and gradual change of $\langle m \rangle$ are shown in the lower panel. The corresponding path of $h(T)$ for each case is also shown in the phase diagram (upper panel). 
(b) Phase diagram of the antiferromagnetic Ising model for two-step SC transitions and typical temperature dependences of $\langle m \rangle$ (lower panel). The red line denotes the critical line for the antiferromagnetic order $\langle m_{\rm AF} \rangle$, in which second-order transitions occur except $T=0$. The dotted line gives $\langle m_{\rm A} \rangle=0$ and $\langle m_{\rm B} \rangle \neq 0$ ($\langle m_{\rm B}\rangle =0$ and $\langle m_{\rm A} \rangle \neq 0$). The critical temperature $T_{\rm c}$ at $h=0$ is given by $T_{\rm c}=z_{\rm AF} |J_{\rm AF}|$ in the mean-field theory. Temperature dependences of $\langle m \rangle$ for a two-step continuous transition (HS to AF to LS) and a one-step continuous transition (HS to AF) are drawn in the lower panel and the corresponding paths of $h(T)$ are given in the upper panel. 
}
\label{phase_diag_MF}
\end{figure}

Using a phase diagram ($T$, $h$) for the Ising model shown in the upper panel of Fig.~\ref{phase_diag_MF} (a), we can obtain a better outlook to discuss the feature of the single-step SC transition. Raising the temperature in the model (\ref{HamWP}) causes a change of the order parameter, $\langle m \rangle=\frac{1}{N} \sum_i \langle \sigma_i \rangle$,  along a line $h(T)$ in the phase diagram of the Ising model. The high spin fraction is $f_{\rm HS}=\frac{\langle m \rangle+1}{2}$. 

We define $T_{\rm c}$ as the critical temperature of the Ising model without the field. In the mean-field theory, $T_{\rm c}=z_{\rm F}J_{\rm F}$, where $z_{\rm F}$ is the coordination number. We also define $T_{\rm cross}(=\frac{D}{{k_{\rm B}}\ln g})$ as the temperature at which the effective field vanishes, i.e., $h(T)=0$. Depending on the relation between $T_{\rm c}$ and $T_{\rm cross}$, the transition between the LS and HS phases is classified into three cases.\cite{Nishino_JCP} 
Typical three cases of temperature dependence of $\langle m \rangle$ are shown in the lower panel and corresponding paths (oblique lines) of $h(T)$ are given in the upper panel. We find (I) gradual change when $T_{\rm c} < T_{\rm cross}$ and  (II) discontinuous (first-order) transition when $T_{\rm c} > T_{\rm cross}$.  The second-order transition (case III) is realized only when $T_{\rm c} = T_{\rm cross}$, which corresponds to the middle case in the lower panel. 
Considering the relation between $h(T)$ and the spinodal lines (blue lines in the upper panel), we can classify the types of transitions in more detail.~\cite{Miya_Proc}

\begin{figure}[h]
  \begin{center}
    \includegraphics[width=61mm]{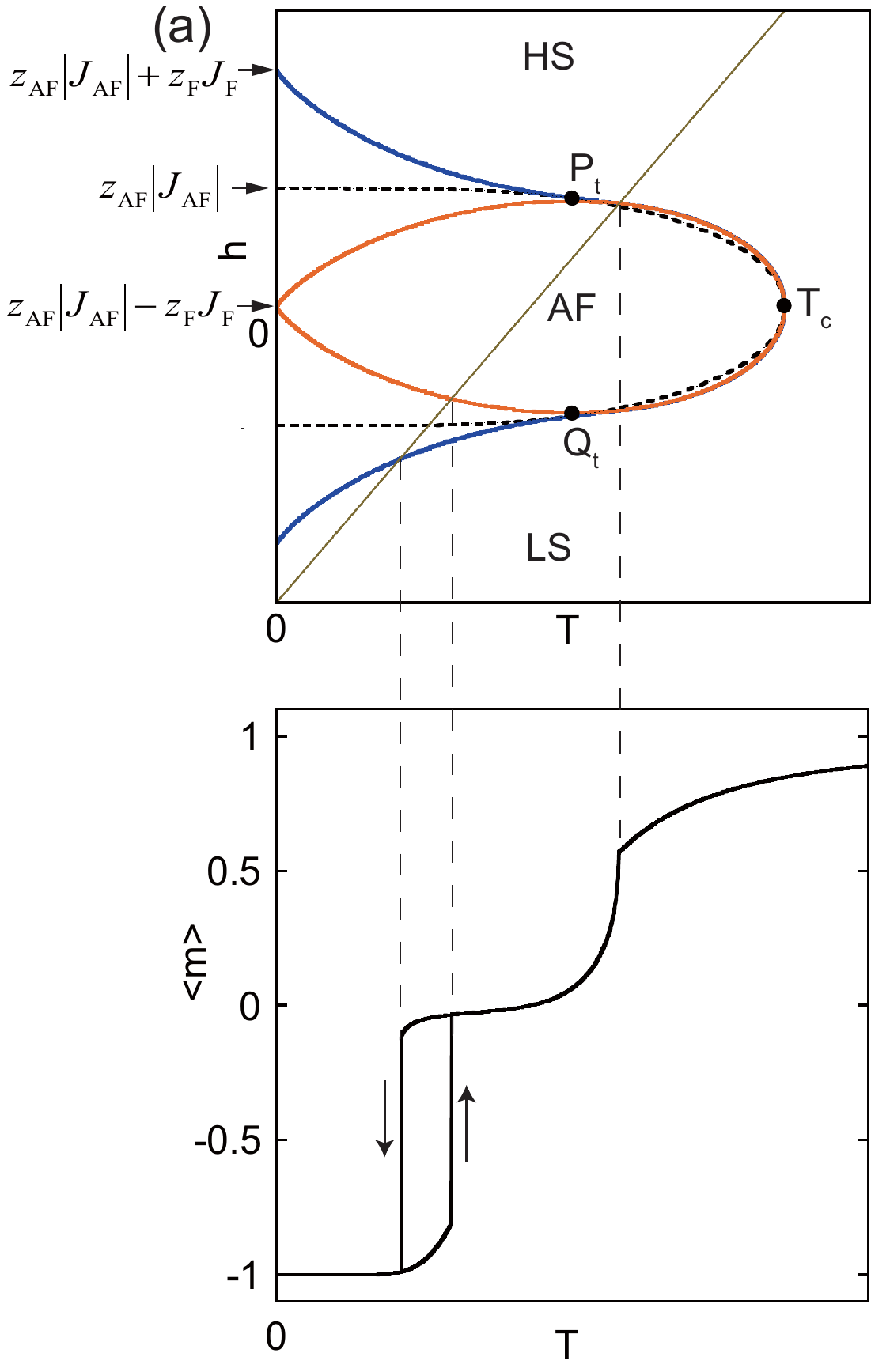}
     \includegraphics[width=48.6mm]{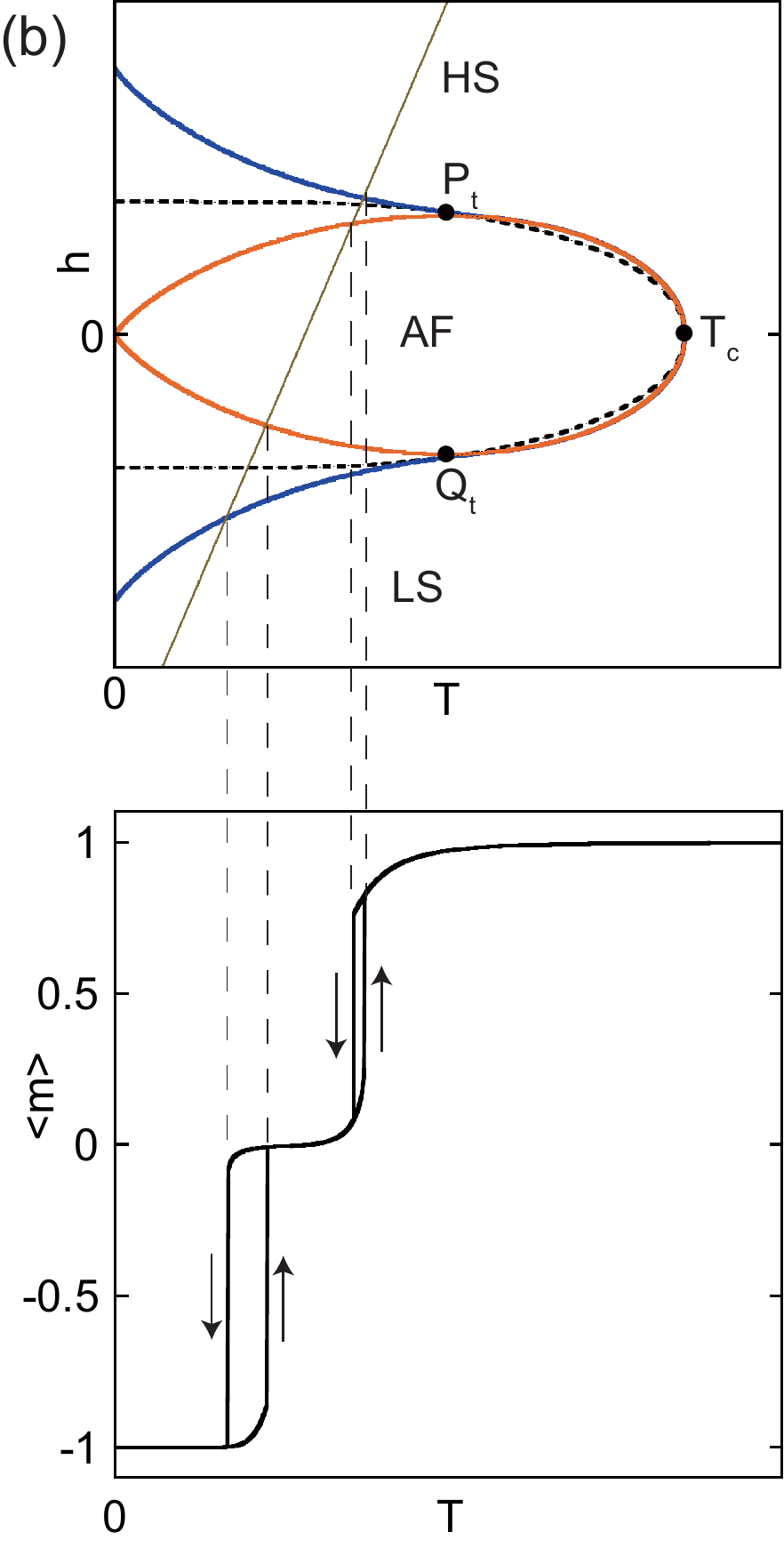}
     \includegraphics[width=49mm]{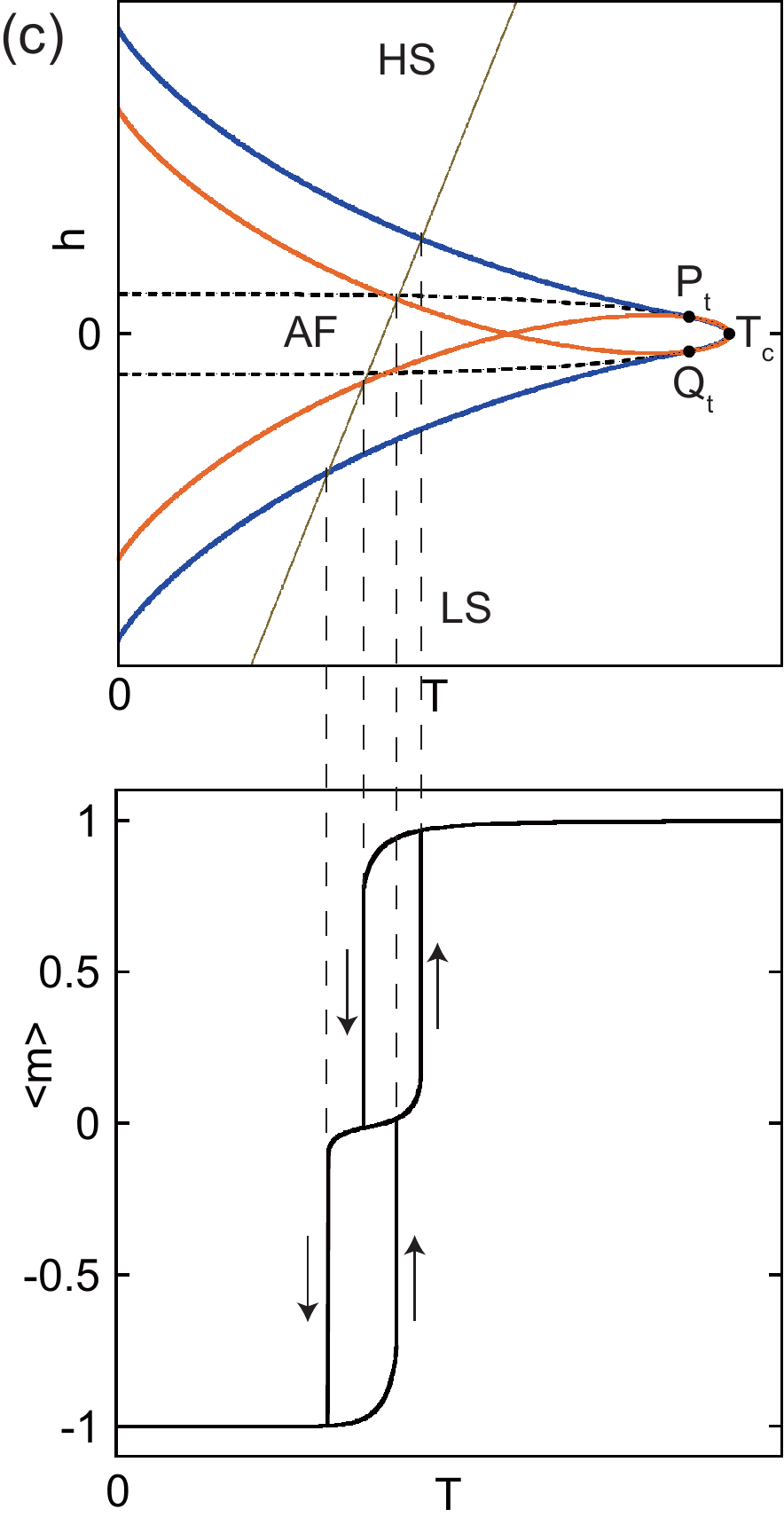}
  \end{center}
\caption{(color online)  (a) Phase diagram of the Ising model with 
antiferromagnetic and ferromagnetic interactions for two-step SC transitions 
(upper panel) and typical temperature dependence of $\langle m \rangle$ (lower panel) of a two-step SC transition with first-order (lower $T$) and second-order (higher $T$) transitions. 
The critical temperature $T_{\rm c}$ at $h=0$ is given by $T_{\rm c}=z_{\rm AF} |J_{\rm AF}|+ z_{\rm F} J_{\rm F}$ in the mean-field theory. Tricritical points are shown by $P_{\rm t}(T_{\rm t},h_{\rm t})$ and $Q_{\rm t}(T_{\rm t},-h_{\rm t})$. The blue lines denote the limit of the metastability of the antiferromagnetic-like phase and the upper (lower) red line for $0 \le T < T_{\rm t}$ denotes the limit of the metastability of the HS (LS) phase. The red line for $ T_{\rm t} \le T  \le T_{\rm c}$ is the critical line for antiferromagnetic-like order $\langle m_{\rm AF} \rangle$. 
(b) A two-step SC transition with double first-order transitions. 
(c) Another type of two-step SC transition with double first-order transitions. Unlike the case (b), when $J_{\rm F}$ is relatively large, the decay temperature of the metastable HS phase can be lower than that of the metastable LS phase. 
 }

\label{phase_diag_two-step}
\end{figure}

Ising-like models have been also extended to study two-step SC transitions and 
several important aspects were successfully clarified.~\cite{
Bousseksou2,Nishino_two-step,Kamel} The Ising-like model for two-step SC in bipartite lattices is given by 
\begin{eqnarray}
{\cal H}=- &&J_{\rm F}\sum_{\langle i\in {\rm A},j\in {\rm A}\rangle }\sigma_{i}^{{\rm A}%
}\sigma_{j}^{{\rm A}}-J_{\rm F}\sum_{\langle i\in {\rm B},j\in {\rm B}\rangle }\sigma_{i}^{%
{\rm B}}\sigma_{j}^{{\rm B}}  \nonumber \\
&-&J_{{\rm AF}}\sum_{\langle i\in {\rm A},j\in {\rm B}\rangle
}\sigma_{i}^{A}\sigma_{j}^{B} -h(T) 
\sum_{i}\left( \sigma_{i}^{A}+\sigma_{i}^{B}\right),
\label{Ham_two-step}
\end{eqnarray}
where $h(T)=-\frac{1}{2}\left(D-{k_{\rm B}T}\ln g\right)$. 
Here A and B denote equivalent sublattices and 
$J_{{\rm F}}$ is the ferromagnetic-like intra-sublattice
interaction and $J_{{\rm AF}}$ is the antiferromagnetic-like 
($J_{{\rm AF}}<0$) inter-sublattice interaction. 

First we review the case of pure antiferromagnetic interaction ($J_{\rm F}=0$). 
We depict a phase diagram ($T$, $h$) of the antiferromagnetic Ising model by the mean-field theory in the upper panel of Fig.~\ref{phase_diag_MF} (b).\cite{note1} 
Two intrinsic order parameters $\langle m_{{\rm A}} \rangle$ and $\langle m_{{\rm B}} \rangle$, associated with the two sublattices A and B, exist, i.e., $\langle m_{{\rm A}} \rangle=\frac{1}{N_{\rm A}} \sum_{i \in {\rm A}} \langle \sigma^{{\rm A}}_i \rangle$ and $\langle m_{{\rm B}}\rangle =\frac{1}{N_{\rm B}}
\sum_{i \in {\rm B}} \langle \sigma^{{\rm B}}_i \rangle$ correspond respectively to the net magnetization per site in the sublattices A and B, respectively. $N_{\rm A}$ ($N_{\rm B}$) is the number of the sites in A (B) sublattice and $N_{\rm A}=N_{\rm B}$. 
We define two order parameters as $\langle m \rangle=\frac{\langle m_{{\rm A}}\rangle+ \langle m_{{\rm B}}\rangle}{2}$ and $\langle m_{\rm AF} \rangle=\frac{\langle m_{{\rm A}}\rangle- \langle m_{{\rm B}}\rangle}{2}$. The former and latter correspond to ferromagnetic-like and antiferromagnetic-like orders.

The red line denotes the critical line in which the antiferromagnetic-like (AF) order $\langle m_{\rm AF} \rangle$ appears and it is the border between the regions of $\langle m_{\rm A} \rangle=\langle m_{\rm B} \rangle$ and $\langle m_{\rm A} \rangle \neq \langle m_{\rm B} \rangle$, which causes second-order phase transitions. Only at $T=0$, the transition is discontinuous. The dotted line gives $\langle m_{\rm A} \rangle=0$ and $\langle m_{\rm B} \rangle \neq 0$ ($\langle m_{\rm B} \rangle=0$ and $\langle m_{\rm A} \rangle \neq 0$). 
 Thus, only continuous (second-order) SC transitions are realized. 
The lower panel of Fig.~\ref{phase_diag_MF} (b) shows examples of a two-step (HS to AF to LS) continuous transition and a one-step (HS to AF) continuous transition, and corresponding paths (oblique lines) of $h(T)$ are also drawn in the upper panel. 

If $J_{\rm F} \neq 0$, the metastable region of the antiferromagnetic-like phase and that of the ferromagnetic-like phase appear at low temperatures and they exist in the region $0 \le T < T_{\rm t}$, where $T_{\rm t}$ is the temperature of tricritical points $P_{\rm t}(T_{\rm t}, h_{\rm t})$ and $Q_{\rm t}(T_{\rm t}, -h_{\rm t})$ (see the upper panel of \ref{phase_diag_two-step} (a)). 
For $0 \le T < T_{\rm t}$, the upper and lower blue lines denote the limit of the metastability of AF phase (i.e., AF spinodal line), while the upper (lower) red line corresponds to the limit of the metastable HS (LS) phase (i.e., HS (LS) spinodal line). The limit of the field for the metastable AF phase at $T=0$ is $h=z_{\rm AF} |J_{\rm AF}|+ z_{\rm F} J_{\rm F} \equiv h_{\rm c}$ and $h=-h_{\rm c}$, and the limit of the field for the metastable HS (LS) phase at $T=0$ is $h=z_{\rm AF} |J_{\rm AF}|- z_{\rm F} J_{\rm F} \equiv h_{\rm 0}$  ($h=-h_{\rm 0}$). 
The critical temperature (Neel point) $T_{\rm c}$ at $h=0$ is given by $T_{\rm c}=z_{\rm AF} |J_{\rm AF}|+ z_{\rm F} J_{\rm F}$  in the mean-field theory. Here $z_{\rm AF}$ ($z_{\rm F}$) is the coordination number for the inter(intra)-sublattice nearest-neighbor sites. 
In the region $T_{\rm t} \le  T    \le T_{\rm c}$, the red line denotes the critical line of the antiferromagnetic-like order $m_{\rm AF}$ for second-order phase transitions as the same as the case $J_{\rm F} = 0$. 

The lower panel of Fig.~\ref{phase_diag_two-step} (a) illustrates an example of a two-step transition with first-order (lower $T$) and second-order (higher $T$) phase transitions. In Fig.\ref{phase_diag_two-step} (b), an example of a two-step transition with double first-order transitions is depicted. 
Depending on the relation between two tricritical points ($P_{\rm t}$ and $Q_{\rm t}$) and the line $h(T)$, the two-step transitions are classified into three cases: (I) first-order and second-order, (II) double first-order, and (III) double second order. 
When the line $h(T)$ locates above the two tricritical points, the type (II) is realized, while $h(T)$ locates below the two tricritical points, the type (III) is realized. 
When $h(T)$ locates between the two tricritical points, the type (I) occurs. 
Figs.\ref{phase_diag_two-step} (a) and (b) correspond to the types (I) and (II), respectively.  

When $J_{\rm F}$ is so large as $z_{\rm F} J_{\rm F} > z_{\rm AF} |J_{\rm AF}|$, another pattern of two-step transitions of double first-order transitions can be realized as depicted in Fig.\ref{phase_diag_two-step} (c). 
 Here the limit of the metastability of the HS (LS) phase (red line) appears at the lower ($h<0$) (upper ($h>0$)) half plane, and thus the decay temperature of the metastable HS phase can be lower than the decay temperature of the metastable LS phase. 
In SC compounds, this case of double first-order transitions has not yet reported in experiments but it may be found in the future. 
The three patterns of two-step transitions were also studied by Monte Carlo methods.\cite{Nishino_two-step}
In the elastic interaction model (Eq.~\ref{H_tot}), the elastic interactions ($k_1$ and $k_2$) and the short-range interaction $J$ take the roles of the interactions $J_{\rm F}$ and $J_{\rm AF}$ of the Ising-like models.

\section{Summary and discussion}
\label{summary}

We obtained a phase diagram as a function of the short-range interaction $J$, where we found that both the ferromagnetic-like and antiferromagnetic-like transition temperatures are enlarged by the elastic interaction. 
However, we found that the natures of ferromagnetic-like and antiferromagnetic-like phase transitions are qualitatively different. 
In the case of the ferromagnetic-like phase transition, the elastic interaction acts as an effective long-range interaction and it significantly enhances the ferromagnetic-like ordering, where the system belongs to the mean-field universality class. 
The increase of the critical temperature is much larger than 
that expected from the Ising interaction. The synergetic effect of the elastic interaction and the short-range interaction amplifies the ferromagnetic-like ordering and causes high critical temperatures. 

In sharp contrast to this case, in the case of the antiferromagnetic-like phase transition, the system belongs to the Ising universality class. In this case the long-rang interaction due to the elastic interaction is irrelevant, and clustering of the ordered phases is observed. 
The elastic interaction raises the critical temperature of antiferromagnetic-like order, as well. This is because the staggered structure is stable for the elastic interaction, but the contribution to enhancement is small. The antiferromagnetic order is mainly enhanced by the short-range interaction. 

We also confirmed different natures in the configurations of the correlation function near the critical point in the cases of ferromagnetic-like and antiferromagnetic-like phase transitions.
Besides, we found that metastable ferromagnetic-like and antiferromagnetic-like phases exist near the region in which both orders are nearly degenerate. 

The present study was performed for the two-dimensional model, but the conclusion can be extended to cases of three dimensions because the physical mechanisms studied in this paper is the same as in three dimensions.

\begin{figure}[t]
  \begin{center}
     \includegraphics[width=130mm]{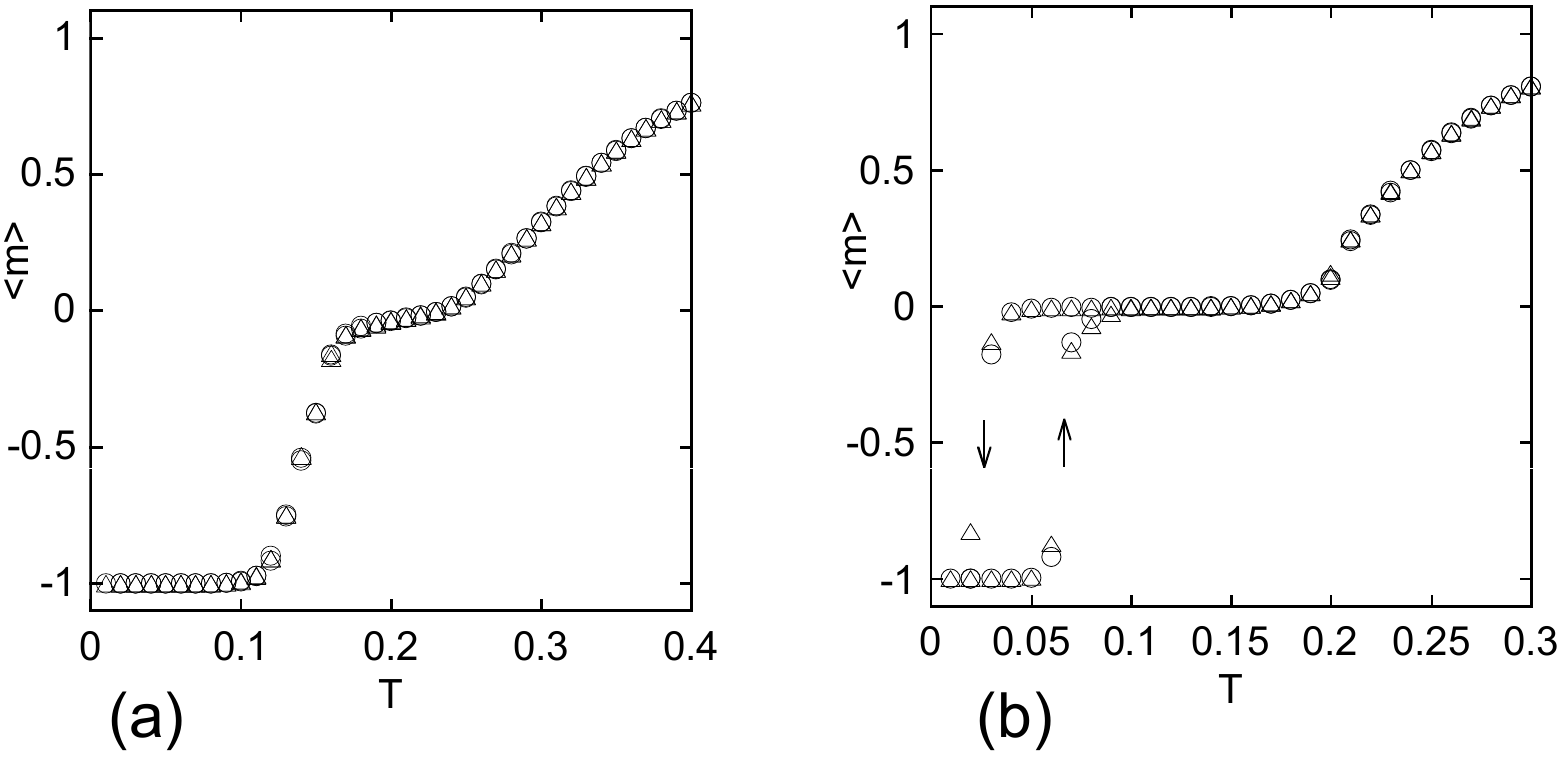}
  \end{center}
\caption{(color online) (a) Two-step SC transition with double continuous (LS $\leftrightarrow$ AF and AF $\leftrightarrow$ HS) transitions. $D=1.6$ and $g=1000$. The symbols $\bigcirc$ and $\triangle$ denote 
$\langle m \rangle$ for $L=20$ and $L=40$, respectively.  
(b) Two-step SC transition with first-order (LS $\leftrightarrow$ AF) and continuous (AF $\leftrightarrow$ HS) transitions. $D=1.02$ and $g=1000$.}
\label{Fig_two-step}
\end{figure}

In the present study, we focused on the critical properties of the model 
along the coexistence line, i.e., ${\cal H}_{\rm eff}=0$. 
In the SC system, ${\cal H}_{\rm eff}$ changes with the temperature, 
and reflecting the structure around the coexistence line, the system 
can exhibit a two-step SC transition with temperature change. 
We depict two types of two-step transitions in Figs.~\ref{Fig_two-step} (a) and  (b). The former shows double continuous transitions and the latter shows first-order (lower $T$) and continuous (higher $T$) transitions. 
In this Monte Carlo simulation, we used 1000,000 MCS to equilibrate and the following 1000,000 MCS to measure $\langle m \rangle$.  
From the results of this study, it is found that the SC phase transition between the HS phase and the intermediate phase (antiferromagnetic-like phase) can be a second order phase transition of the short-range Ising interaction type, while the SC phase transition between the HS and LS phases is of the mean-field type when it is of the second order. The approach of the present study enables us to capture new aspects of two-step SC transitions from the view point of the elastic interaction and the short-range interaction. 

The phase diagrams of HS, LS, and AF phases for the elastic interaction model with the short-range interaction may have similar features to those obtained by Ising-like models in the mean-field theory in Sec.~\ref{sec_Ising-like} (see Fig.~\ref{phase_diag_MF} and Fig.~\ref{phase_diag_two-step}). 
However, the nature of the phase transition of the model is not the same as that of the Ising-like models. The universality classes of the transitions are different from those of the Ising-like models.  
The regions of the metastability for the HS, LS, and AF phases are determined by local interactions in the Ising-like models, while those are determined by both local and global stability in the elastic model. Thus the quantitative features of the phase diagrams in the elastic model are different from those in the Ising-like models, which is non-trivial. The details of the phase diagrams and correspondence to SC compounds will be studied in the future.\cite{Nishino_prep}

When $J=J_0$, the antiferromagnetic-like phase appears although the ground 
states ($T=0$) of the two phases are degenerate at this value of $J$.  
This fact indicates that the antiferromagnetic-like phase is favorable at finite temperatures. In the text we considered a reason why the antiferromagnetic-like phase is realized  easier than the ferromagnetic-like phase. 
We may attribute it to a dynamical effect. The average fraction of HS and LS molecules of this ordered phase is almost the same as in the disordered phase and the antiferromagnetic-like phase can be realized without modification of the density. We may also consider this fact from the view point of entropy, i.e., a kind of mechanism of order by disorder. The detailed study will be done in the future. 

\section*{Acknowledgments}
The authors thank Professor Fran\c cois Varret for pointing out the fact 
that the elastic interaction also favors the antiferromagnetic-like configuration. The present work was supported by the Mitsubishi foundation, 
Grant-in-Aid for Scientific Research on Priority Areas, KAKENHI (C)  23540381, 
and also the Next Generation 
Super Computer Project, Nanoscience Program from MEXT of Japan. 
The numerical calculations were supported by the supercomputer center of
ISSP of University of Tokyo.

\clearpage

\appendix 

\section{Interactions to maintain the square symmetry and the coexistence of high-spin and low-spin phases }
\label{appendix_angle}
\begin{figure}[t]
  \begin{center}
     \includegraphics[width=40mm]{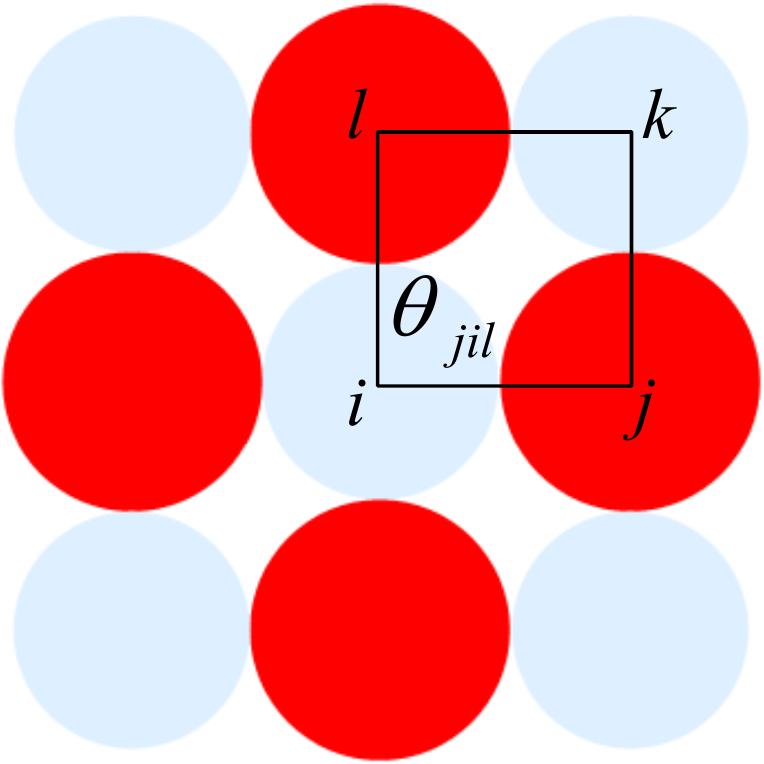}
  \end{center}
\caption{(color online) The definition of $\theta_{jil}$. 
}
\label{angle}
\end{figure}

If only the nearest neighbor interaction (Eq.~(\ref{eq_H})) is used, the lattice may distort into a rhombic shape.  
To prohibit such deformation, we need some additional interaction.
In the text, we applied the next nearest interaction ${\cal H}_{\rm nnn}$ 
(Eq. (\ref{eq_Hnnn})). 
We may consider other interactions instead of the choice of ${\cal H}_{\rm nnn}$. A kind of interactions to maintain the angle between bonds for desired lattices has been frequently used.\cite{Dunweg,Laradji,Zhu} Thus in the same way we focus on the angle ($\theta_{jil}$) between the bonds $i-j$ and $i-l$ (Fig.~\ref{angle}), which is defined by the relation 
\begin{equation}
\cos \theta_{jil}=\frac{ \vec{r}_{i,j} \cdot \vec{r}_{i,l}}{r_{i,j}  r_{i,l}}. 
\end{equation}
To maintain a square lattice, i.e., $\theta_{jil}$ equal to $\pi/2$, the following interaction can be adopted. 
\begin{equation}
{\cal H_\theta}= C \sum \cos^2 \theta_{jil},  
\label{angle_model}
\end{equation}
where $C$ is a positive constant, and the summation is taken over all pairs of bonds. 

For the configuration of the ferromagnetic-like phase or antiferromagnetic-like phase in Fig.~\ref{Fig_conf_T0}, ${\cal H_\theta}$ takes the minimum value 
${\cal H_\theta} = 0$, 
and  this term plays a role in maintaining the square symmetry. 
In this interaction, the ferromagnetic-like and antiferromagnetic-like phases have the same energies at $T=0$ and both phases are the ground state. 
The origin of $J$ is zero ($J_0=0$) in this case.

Unlike the case of Eq. (\ref{eq_Hnnn}), in this choice (Eq. (\ref{angle_model})), the configuration entropy is different for HS molecules and LS molecules, as analyzed below. 
We consider the motion of molecules around the position in the complete LS (HS) phase (as in Figs~\ref{Fig_conf_T0} (a) and (b)).
Let the fluctuation of a molecule in $x$-direction ($y$-direction) defined as $dx$ ($dy$). 
The potential term for the molecule $U$, given by $\cos^2 \theta_{jil}$, has the form  
\beq
U \propto {(dx+dy)^2\over R^2}
\label{eqU}
\eeq 
in the leading term of $dx$ and $dy$, where $R$ is the molecular radius. 
Because $R$ is larger in the HS state and the entropy gain is larger in the HS state, the HS state is more favorable than the LS state. 
Indeed, we observed that only HS state is realized in the simulation. 
LS and HS states are not symmetric anymore. This situation is similar to the Ising model with nonzero magnetic field. Thus critical phenomena do not occur in this case.  
A field for cancelation of this difference was artificially applied when critical properties were studied.~\cite{Dunweg,Laradji,Zhu} 
Thus, we adopt the present choice to avoid this complication, where ${\cal H}_{\rm nnn}$ has no $R$-dependence as $dx^2+dy^2$ in the leading term and the system has a critical point. 
Rigorously speaking, the HS state is more favorable than the LS state even in this case. However, ${\cal H}_{\rm nnn}$ has $R$-dependence at higher order and 
the difference is small and can be ignored in the practical calculation (no influence on simulations). This analysis holds in three dimensions. Namely, Eq. (\ref{eqU}) has the dependence of ${(dx+dy)^2\over R^2}$, ${(dy+dz)^2\over R^2}$, or ${(dz+dx)^2\over R^2}$ for the fluctuation ($dx$, $dy$, $dz$), while Eq.~(\ref{eq_Hnnn}) has no $R$-dependence as $dx^2+dy^2$, $dy^2+dz^2$, or $dz^2+dx^2$.  

In order to confirm the above mentioned effects, we performed Monte Carlo simulations with the interaction (\ref{angle_model}) with $C=4$.
For $J<0$ we found that antiferromagnetic-like transition occurs and it has the same critical properties as in the case of the text. Namely, curves of $T$ dependence of $U_4^{\rm AF}$ cross for different system sizes $L$ and the value of $U_4^{\rm AF}$ at the crossing point is $U_4^{\rm AF}=0.61$ (not shown). 
For example, $T_{\rm c}$ is estimated $T_{\rm c}=0.138$ when $J=-0.04$. 
It is clear that the interaction (\ref{angle_model}) has the same contribution to the two coexistent anitiferromagnetic-like phases and has no effect on the critical properties.   

\begin{figure}[t]
  \begin{center}
     \includegraphics[width=130mm]{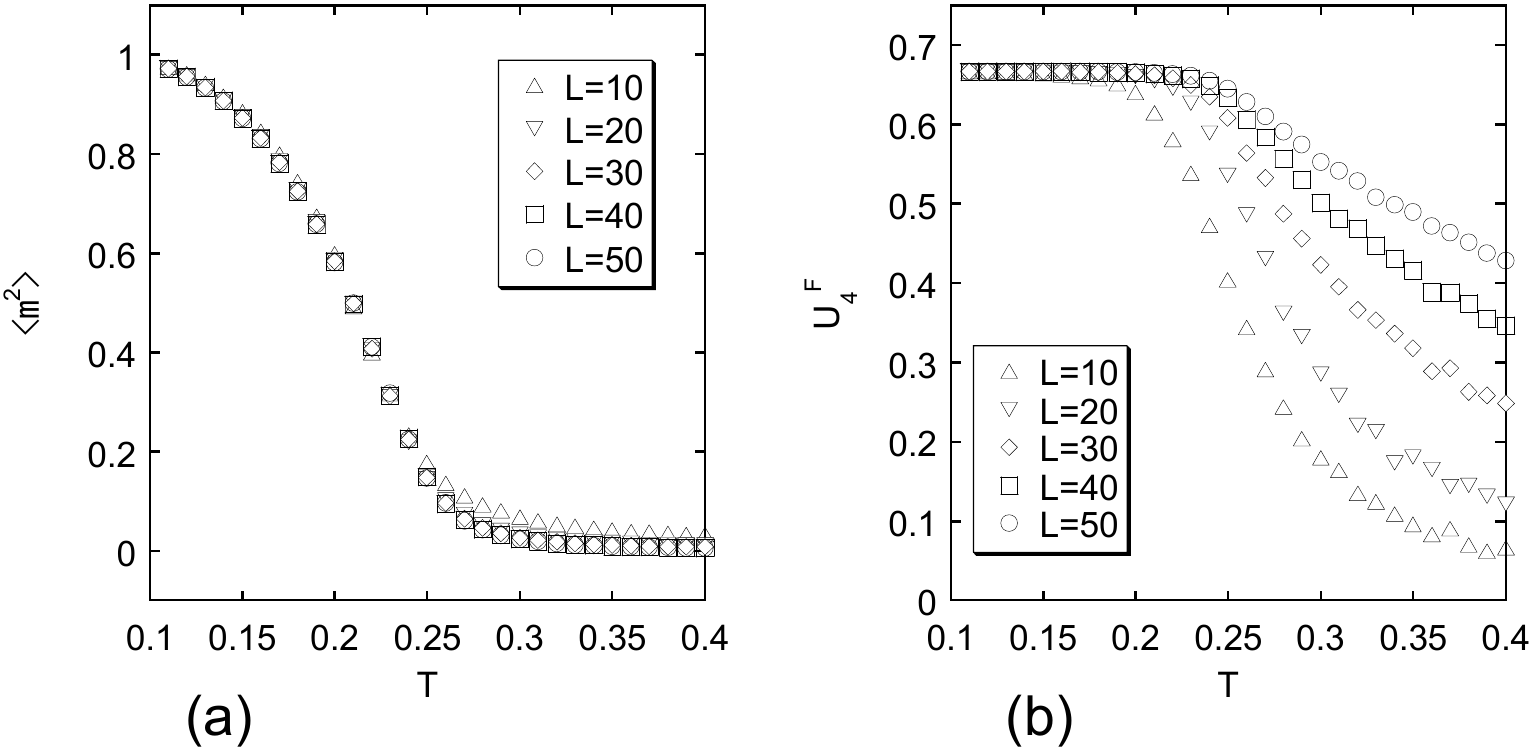}
  \end{center}
\caption{(color online) 
Temperature dependence of $\langle m^2  \rangle$ (a) and temperature dependence of $U^{\rm F}_4$ (b) for various system sizes $L$ when $J=0.04$. }
\label{angle_binder}
\end{figure}

In contrast, for the ferromagnetic-like region, it has a big effect on the critical property. In Fig.~\ref{angle_binder} (a) the temperature dependence of $\langle m^2 \rangle$ is given for $J=0.04$. Here we find a sharp change of magnetization $\langle m^2 \rangle$ which indicates ferromagnetic-like phase transitions, and the magnetization seems to have a critical point. 
 However, if we plot the Binder cumulant for the ferromagnetic-like transition, a strange dependence on the size was found.  
In contrast to the antiferromagnetic-like case, curves of $T$ vs. $U_4^{\rm AF}$ for different system sizes do not cross in this ferromagnetic-like case, shown in Fig.~\ref{angle_binder} (b). Thus we conclude that the critical point does not exist in this case. 
When $J=0.001$, an antiferromagnetic-like transition takes place, and for $J \geq 0.01$ the system shows a ferromagnetic-like transition. There is a critical value of $J$ ($J_{\rm c}$) between the ferromagnetic-like and antiferromagnetic-like phases. This critical value locates in the region of $0.001<J_{\rm c}<0.01$. This is considered a similar finite temperature effect which is discussed in the text. 

As another choice,
we may adopt a next-nearest neighbor interaction to realize $J_0=0$. 
\begin{equation}
{\cal H}_{\rm nnn}={k_2 \over 2}\sum _{\langle\langle i,k \rangle\rangle}[r_{i,k}-2\sqrt{2}\bar{R}_{ijkl}]^2.
\label{eq_nnn}
\end{equation}
Here the next nearest neighbor interaction $i-k$ depends on the bonds $i-j$, $j-k$, $k-l$ and $l-i$, and  
we define 
\begin{eqnarray}
2\bar{R}_{ijkl} &=& \big[\big(R_i(\sigma_i)+R_j(\sigma_j)\big)+\big(R_j(\sigma_j)+R_k(\sigma_k)\big)+\big(R_k(\sigma_k)+R_l(\sigma_l)\big) \\ \nonumber  
&&+\big(R_l(\sigma_l) +R_i(\sigma_i)\big) \big]/4 \\ \nonumber  
             &=& 
\big[ R_i(\sigma_i)+R_j(\sigma_j)+R_k(\sigma_k)+R_l(\sigma_l) \big] /2.
\end{eqnarray}
In this case the states of Fig.~\ref{Fig_conf_T0} (ferromagnetic-like and antiferromagnetic-like configurations) are the ground state and $J_0=0$. However, this model gives a similar situation of non-crossing of $U^{\rm F}_4$-$T$ curves although some cases of $L$ show crossing. 
This reason is not the same as the case (\ref{angle_model}). The reason is not so clear but 4-body interactions may cause such complex dependences. 
Thus, although $J_0\neq 0$, we adopted Eq. (\ref{eq_Hnnn}) for the purpose of maintaining the square lattice in the present work.

\end{document}